\documentclass[12pt]{article}
\usepackage{epsfig}

\makeatletter 
\@addtoreset{equation}{section} 
\makeatother 
\renewcommand{\theequation}{\thesection.\arabic{equation}} 

\def\draftlabel#1{{\@bsphack\if@filesw {\let\thepage\relax
   \xdef\@gtempa{\write\@auxout{\string
      \newlabel{#1}{{\@currentlabel}{\thepage}}}}}\@gtempa
   \if@nobreak \ifvmode\nobreak\fi\fi\fi\@esphack}
        \gdef\@eqnlabel{#1}}
\def\@eqnlabel{}
\def\@vacuum{}
\def\draftmarginnote#1{\marginpar{\raggedright\scriptsize\tt#1}}

\def\numberbysection{\@addtoreset{equation}{section}
        \def\theequation{\thesection.\arabic{equation}}}

\def\titlepage{\@restonecolfalse\if@twocolumn\@restonecoltrue\onecolumn
     \else \newpage \fi \thispagestyle{empty}\c@page\z@
        \def\thefootnote{\fnsymbol{footnote}} }

\def\endtitlepage{\if@restonecol\twocolumn \else \newpage \fi
        \def\thefootnote{\arabic{footnote}}
        \setcounter{footnote}{0}}  

\catcode`@=12
\relax

\def\figcap{\section*{Figure Captions\markboth
        {FIGURECAPTIONS}{FIGURECAPTIONS}}\list
        {Figure \arabic{enumi}:\hfill}{\settowidth\labelwidth{Figure
999:}
        \leftmargin\labelwidth
        \advance\leftmargin\labelsep\usecounter{enumi}}}
 \relax

\def\draftlabel#1{{\@bsphack\if@filesw {\let\thepage\relax
   \xdef\@gtempa{\write\@auxout{\string
      \newlabel{#1}{{\@currentlabel}{\thepage}}}}}\@gtempa
   \if@nobreak \ifvmode\nobreak\fi\fi\fi\@esphack}
        \gdef\@eqnlabel{#1}}
\def\@eqnlabel{}
\def\@vacuum{}
\def\draftmarginnote#1{\marginpar{\raggedright\scriptsize\tt#1}}

\def\draft{\oddsidemargin -.5truein
        \def\@oddfoot{\sl preliminary draft \hfil
        \rm\thepage\hfil\sl\today\quad\militarytime}
        \let\@evenfoot\@oddfoot \overfullrule 3pt
        \let\label=\draftlabel
        \let\marginnote=\draftmarginnote
   \def\@eqnnum{(\theequation)\rlap{\kern\marginparsep\tt\@eqnlabel}%
\global\let\@eqnlabel\@vacuum}  }

\def\be{\begin{equation}}
\def\ee{\end{equation}}
\def\ba{\begin{eqnarray}}
\def\ea{\end{eqnarray}}

\newcommand{\beq}{\begin{equation}}
\newcommand{\eeq}[1]{\label{#1}\end{equation}}
\newcommand{\ber}{\begin{eqnarray}}
\newcommand{\eer}[1]{\label{#1}\end{eqnarray}}
\newcommand{\eqn}[1]{(\ref{#1})}

\def\tr{\,{\rm tr}\,}

\def\a{\alpha}
\def\b{\beta}
\def\g{\gamma}
\def\G{\Gamma}
\def\d{\partial}
\def\e{\epsilon}
\def\p{\psi} 

\def\cb{{\overline\chi}}

\def\m{\mu}
\def\n{\nu}
\def\r{\rho}
\def\l{\lambda}
\def\lb{{\overline\lambda}}

\def\s{\sigma}

\def\f{\phi}
\def\vf{\varphi}

\def\ks{{k \kern-.5em /}}
\def\es{{\e \kern-.4em /}}
\def\ds{{\partial \kern-.5em /}}
\def\Ds{{D \kern-.6em /}}

\def\half{{\textstyle {1\over 2}}}

\def \ha {{1\over 2}}
\def\inv{^{\raise.15ex\hbox{${\scriptscriptstyle -}$}\kern-.05em 1}}

\def\wt{\widetilde}

 \def\PL #1 #2 #3 {Phys.~Lett.~{\bf #1} (#2) #3}
 \def\NP #1 #2 #3 {Nucl.~Phys.~{\bf #1} (#2) #3}
 \def\PR #1 #2 #3 {Phys.~Rev.~{\bf #1} (#2) #3}
 \def\PRL #1 #2 #3 {Phys.~Rev.~Lett.~{\bf #1} (#2) #3}
 \def\CMP #1 #2 #3 {Comm.~Math.~Phys.~{\bf #1} (#2) #3}
 \def\IJMP #1 #2 #3 {Int.~J.~Mod.~Phys.~{\bf #1} (#2) #3}
 \def\JETP #1 #2 #3 {Sov.~Phys.~JETP.~{\bf #1} (#2) #3}
 \def\PRS #1 #2 #3 {Proc.~Roy.~Soc.~{\bf #1} (#2) #3}
\def\IJMP #1 #2 #3 {Int.~J.~Mod.~Phys.~{\bf #1} (#2) #3}
 \def\AP #1 #2 #3 {Ann.~Phys.~{\bf #1} (#2) #3}
 \def\MPL #1 #2 #3 {Mod.~Phys.~Lett.~{\bf #1} (#2) #3}

\def\a{\alpha}
\def\adot{{\dot\alpha}}
\def\b{\beta}
\def\m{\mu}
\def\s{\sigma}
\def\n{\nu}
\def\r{\rho}
\def\l{\lambda}
\def\lb{{\bar\lambda}}
\def\L{\Lambda}
\def\g{\gamma}
\def\G{\Gamma}
\def\t{\theta}
\def\tb{{\bar\theta}}
\def\sm{\sigma^\m}
\def\smb{\bar\sigma^\m}
\def\d{\partial}
\def\rmd{{\rm d}}
\def\vf{\varphi}
\def\f{\phi}
\def\F{\Phi}
\def\fd{\phi_D}
\def\Fd{\Phi_D}
\def\ix{\int {\rm d}^4 x \, }
\def\dt{{\rm d}^2 \t \, }
\def\dtb{{\rm d}^2 \tb \, }
\def\dtt{{\rm d}^2 {\tilde\t} \, }
\def\P{\Psi}
\def\cf{{\cal F}}
\def\cfd{{\cal F}_D}
\def\ad{a_D}
\def\e{\epsilon}
\def\rmd{{\rm d}}
\def\rd{\sqrt{2}}
\def\la{\langle}
\def\ra{\rangle}
\def\vac{\vert 0\rangle}
\def\P{\Psi}
\def\p{\psi}
\def\pb{{\bar \psi}}
\def\dd #1 #2{{\delta #1\over \delta #2}}
\def\tr{{\rm tr}\ }
\def\til{\widetilde}
\def\nm{\nabla_\m}
\def\fmn{F_{\m\n}}
\def\fmnt{{\tilde F}_{\m\n}}
\def\im{{\rm Im}\, }
\def\M{{\cal M}}

\def\crbig{\\\noalign{\vspace {3mm}}}

\def\vac{|0\rangle}

\def\tr{{\rm tr}\, }
\def\ha{{1\over 2}}
\def\wt{\widetilde}
\def\ra{\rangle}
\def\la{\langle}
\def\ov{\overline}
\def\a{\alpha}
\def\b{\beta}
\def\g{\gamma}
\def\G{\Gamma}
\def\e{\epsilon}
\def\f{\phi}
\def\fb{{\ov \phi}}
\def\vf{\varphi}
\def\m{\mu}
\def\n{\nu}
\def\r{\rho}

\def\s{\sigma}
\def\sb{\ov\sigma}
\def\l{\lambda}
\def\L{\Lambda}
\def\p{\psi}
\def\pb{\ov\psi}
\def\cb{\ov\chi}
\def\d{\partial}
\def\dag{\dagger}

\def\da{{\dot\alpha}}
\def\db{{\dot\beta}}

\def\dd{{\dot\delta}}
\def\t{\theta}
\def\tb{{\ov \theta}}
\def\lb{{\ov \lambda}}
\def\eb{{\ov \epsilon}}

\def\Db{\ov D}
\def\M{{\cal M}}
\def\rd{\sqrt{2}}
\def\Tr{{\rm Tr\, }}
\def\F{{\cal F}}
\def\Dint{\int\, {\rm d}^2\theta {\rm d}^2\overline\theta\ }
\def\Fint{\int\, {\rm d}^2\theta\ }
\def\Fbarint{\int\, {\rm d}^2\overline\theta\ }
\def\xint{\int\, {\rm d}^4 x}

\begin{document}
\newcommand{\fig}[3]{\epsfxsize=#1\epsfysize=#2\epsfbox{#3}}



\pagestyle{empty}
\begin{flushright}
\small
NEIP-01-006\\
IHP-2000/32\\
{\tt hep-th/0106246}\\
June 2001
\normalsize
\end{flushright}

\begin{center}


\vspace{.7cm}

{\Large {\bf DUALITY IN ${\cal N}=2$ SUSY GAUGE THEORIES:\break

             low-energy effective action and BPS spectra}}\footnote
             {Lectures given at the Institut Henri Poincar\'e program
             ``Strings, Supergravity and M-theory", January 2001}

\vspace{2.cm}

{\large Adel~Bilal}

\vskip 0.4truecm

Institute of Physics, University of Neuch\^atel\\
   rue Breguet 1, 2000 Neuch\^atel, Switzerland\\
   {\tt adel.bilal@unine.ch}

\vskip 3.0cm


{\bf Abstract}

\end{center}

\begin{quotation}

\small
\noindent
After an introduction to ${\cal N}=2$ susy Yang-Mills theories, I review 
in some detail, for the ${\rm SU}(2)$ gauge group, how the low-energy 
effective action is obtained using
duality and the constraints arising from the supersymmetry.
Then I discuss how knowledge of this action, duality and certain discrete 
symmetries allow us to determine the spectra of stable BPS states at any 
point in moduli space. This is done for gauge group ${\rm SU}(2)$, 
without and with fundamental matter hypermultiplets which may have 
non-vanishing bare masses. In the latter case non-trivial four-dimensional 
CFTs arise at Argyres-Douglas type points.

\end{quotation}

\newpage


\pagestyle{plain}

\setcounter{section}{0}
\section{Introduction\label{Intro}}

The central tool in modern string and M-theory certainly is duality.
Duality has a long history, but it was only since the ground-breaking 
work by Seiberg and Witten in 1994 \cite{SW, SWII} that it has become a useful tool 
and maybe even more: an organising principle and underlying symmetry of
string/M-theory. Dualities were discovered and suspected in string 
theory even before, but it is certainly fair to say that it was in the 
context of ${\cal N}=2$ supersymmetric Yang-Mills quantum field theories
that their power was impressively revealed.

The prehistory of duality probably goes back to Dirac who observed that the
source-free Maxwell equations are symmetric under the exchange of the
electric and magnetic fields. More precisely, the symmetry is $E\to B,\ B\to
-E$, or $\fmn\to \fmnt=\half
\e_{\m\n}^{\phantom{\m\n}\rho\sigma}F_{\rho\sigma}$.\footnote
{$\e_{\m\n\rho\sigma}$ is the flat-space antisymmetric $\e$-tensor with
$\e^{0123}=+1$ and we use $\eta_{\m\n}$ with signature $(1,-1,-1,-1)$.
} 
To maintain
this symmetry in the presence of sources, Dirac introduced, somewhat ad hoc,
magnetic monopoles with magnetic charges $q_m$ in addition to the electric
charges $q_e$, and showed that consistency of the quantum theory requires a
charge quantization condition $q_mq_e=2\pi n$ with integer $n$. Hence the
minimal charges obey $q_m={2\pi\over q_e}$. Duality exchanges $q_e$ and
$q_m$, i.e. $q_e$ and ${2\pi\over q_e}$. Now recall that the electric charge
$q_e$ also is the coupling constant. So duality exchanges the coupling
constant with its inverse (up to the factor of $2\pi$), hence exchanging
strong and weak coupling. This is the reason why physicists are 
so much interested in
duality: the hope is to learn about strong-coupling physics from the
weak-coupling physics of a dual formulation of the theory. Of course, in
classical Maxwell theory we know all we may want to know, but this is no
longer true in quantum electrodynamics.

Actually, quantum electrodynamics is not a good candidate for exhibiting a
duality symmetry since there are no magnetic monopoles, but the latter
naturally appear as solitons in spontaneously broken non-abelian gauge theories
\cite{THO}. Unfortunately, electric-magnetic duality in its simplest form cannot
be a symmetry of the quantum theory due to the running of the coupling
constant (among other reasons). Indeed, if duality exchanges $g^2(\L)
\leftrightarrow {1\over g^2(\L)}$ at some
scale $\L$, in general this won't be true at another scale. 
There are two ways out: either 
the coupling does not run, i.e. if the $\b$-function vanishes as is the case
in certain ($N=4$) supersymmetric extensions of the Yang-Mills theory,
or, if $\b\ne 0$, we can have instead 
$g^2(\L)\leftrightarrow {1\over g^2(\L_D)}$
where $\L_D=\L_D(\L)$ is some dual scale which depends on the 
initial scale $\L$ just in the right way to make the duality possible.
The first possibility led Montonen and Olive \cite{MO} to conjecture 
that duality might be an exact symmetry of $N=4$ susy Yang-Mills 
theory. A nice review of these ideas can
be found in \cite{OLI}. It is the second possibility that is realised for the 
low-energy effective action of ${\cal N}=2$ susy Yang-Mills theory, and 
it is this second case which will occupy the present review.

Let me insist that ${\cal N}=2$ susy Yang-Mills theory is {\it not} duality 
invariant. For simplicity we will restrict to an ${\rm SU}(2)$ gauge group, 
although other gauge groups can be discussed along the same lines. This 
${\rm SU}(2)$ is spontaneously broken down to ${\rm U}(1)$, so that the 
gauge bosons are a massless ``photon" and two massive ``W-bosons" along 
with their superpartners. Since the ``W-bosons" are heavy they can be 
integrated out along with all other heavy states when determining the 
low-energy effective action for the massless ``photon"-multiplet. We 
refer to the underlying ${\rm SU}(2)$ theory as the microscopic theory 
since it governs the UV behaviour, while the theory of the massless 
${\rm U}(1)$ degrees of freedom governs the effective IR dynamics. It 
is this latter theory in which duality is realised as follows. This 
low-energy effective action will turn out to have many gauge-inequivalent 
vacua determined by the scalar (``Higgs") field expectation value, i.e. 
there is a moduli space. Let $g(\m)$ be the effective coupling  where the 
scale $\m$ is given by the scalar vev. The duality then is
\be\label{ui}
 g^2(\m)\ \leftrightarrow\ {1\over g^2(\m_D)}
\ee

What is duality good for ? If it exchanges a coupling $g$ with its inverse, 
it will map weak coupling, say $g^2< 0.1$ to very strong coupling, 
say $g^2> 10$, and vice versa. But a coupling close to unity then again is
mapped to a coupling close to unity. While often this is the more
interesting case, it seems that precisely here duality might be useless.
In fact, this is not so. As an illustration,
let me recall that a somewhat similar duality symmetry appears 
in the two-dimensional
Ising model where it exchanges the temperature with a dual temperature, thereby
exchanging high and low temperature analogous to strong and weak coupling.
It is useful to consider this example in slightly more detail:

\vskip 3.mm
\noindent
\underline{Kramers-Wannier duality of the 2D Ising model}

\noindent
Consider the Ising model on a square lattice with $N$ sites. The
partition function is
\be\label{uii} 
Z=\sum_{\s_l=\pm 1} \exp \left( \b \sum_{<i,j>} J \s_i\s_j \right) \ ,
\ee
where  $\b=1/T$. This can be rewritten as
\be\label{uiii}
 Z(\b) \equiv Z_{\rm HT}(\b)=\sum_{\s_l=\pm 1} \prod_{<i,j>}
\cosh\b J \left[ 1+\s_i\s_j\tanh\b J\right]
\ee
A little thinking shows that this yields the sum of all closed polygons 
${\cal P}$
drawn on the lattice (a polygon may have several disconnected pieces), 
each polygon having a weight factor 
$(\tanh\b J)^{L({\cal P})}$, where $L({\cal P})$ is the length of the 
polygon in lattice units:
\be\label{uiv}
 Z(\b)=2^N \left( \cosh\b J \right)^{2N}
\sum_{{\cal P}} (\tanh\b J)^{L({\cal P})}
\ee
This is an appropriate expansion if $\tanh\b J$ is small, i.e. at high 
temperatures $T=1/\b$.
At low temperature instead, there will be domains of spins pointing 
in the same direction, separated from each other by domain boundaries 
which are polygons ${\cal P}$ drawn on the dual lattice. For a square 
lattice the 
dual lattice is again the same square lattice (only displaced). The energy 
of such a configuration is is obtained from the ground state energy $E_0=-2NJ$
by adding the contributions due to the domain boundaries:
\be\label{uv}
E({\cal P})=E_0 + 2 J L({\cal P})
\ee
so that the appropriate expansion of the partition function at 
low temperature is
\be\label{uvi}
Z(\b) \equiv Z_{\rm LT}(\b)
=e^{-\b E_0} \sum_{{\cal P}} \left(e^{-2\b J}\right)^{L({\cal P})}
\ee
So there are two different expansions of the {\it same} Ising partition 
function. Now define a dual temperature by
\be\label{uvii}
e^{-2\b_D J}=\tanh \b J  \ \Leftrightarrow   e^{-2\b J}=\tanh \b_D J
\ee
so that we can write using first \eqn{uvi} and then \eqn{uiv}
\be\label{uviii}
Z(\b_D)= e^{-\b_D E_0} \sum_{{\cal P}} (\tanh\b J)^{L({\cal P})}
=2^{-N} e^{-\b_D E_0}\left( \cosh\b J \right)^{-2N} Z(\b)\ .
\ee
This is an amazing functional relation for the partition function $Z$ which 
allows the determination of the critical temperature as follows:
if there is a (single) phase transition at some critical value $\b^*$
the free energy $F(\b)=-{1\over \b} \log Z(\b)$ must be singular at $\b=\b^*$. 
The prefactors on the r.h.s. of  \eqn{uviii} will not change the singular 
behaviour of $F$, so we conclude from \eqn{uviii} that if $\log  Z(\b)$ is 
singular, so must be $\log  Z(\b_D)$. But this implies that $\b_D$ must 
also be at the critical value: $\b=\b^*=\b_D$, i.e. the critical temperature 
is the self-dual point. It is given by the solution of
\be\label{uix}
e^{-2\b^* J}=\tanh \b^* J \ \  \Rightarrow \ \
\b^* J = {1\over 2} \log (\sqrt{2}+1) \ .
\ee
We see that for the Ising model, the sole existence of the duality symmetry 
leads to the 
exact determination of the critical temperature as the self-dual point.
Historically this preceeded Onsager's
exact solution by a few years. One may
view the existence of this self-dual point as the requirement that the dual
high and low temperature regimes can be consistently ``glued" together.

Note that the functional relation  \eqn{uviii} not only gives the 
critical temperature but also allows us to obtain quite some 
information on the form of the partition function at any
$N$ and $T$ as follows. We let $x=e^{-2\b J}$ so that 
$e^{-2\b_D J}= {1-x\over 1+x}$. If we
define $Z(\b)\equiv z(x)^N$ then  \eqn{uviii} takes the form 
$z\left(  {1-x\over 1+x}\right) = {2x\over 1-x^2} z(x)$. 
Substituting $z(x)=(1-x) f(\xi)$ with 
$\xi=\log {x+1\over \sqrt{2}}$ this simply becomes 
$f(-\xi)=f(\xi)$ so that $f$ is an even
function of $\xi$, and the self-dual point is $\xi^*=0$. 
Although this is not enough to completely determine the partition 
function, it gives valuable information relating the high
and low-temperature behaviour.

Similarly, in the Seiberg-Witten theory, duality relates the behaviour 
of the effective action at strong coupling to its behaviour at weak 
coupling. Here however, as will be explained below, the requirement
that both regimes can be consistently glued together is much stronger.
While in the Ising model the ``gluing" resulted in the determination of
the critical point, in the ${\cal N}=2$ susy gauge theories we have a
holomorphicity requirement allowing for both regimes to be smoothly
glued together. Adding some information on the asymptotic behaviours at
weak and strong coupling then completely determines the full effective
action for the light fields at any coupling.

\vskip 3.mm
\noindent
\underline{Outline of the paper}

\noindent
Let me give an overview of the material covered in these lectures. 
Since ${\cal N}=2$ supersymmetry plays a  central role, I will spend 
some time and 
space
in the next section to review those notions of supersymmetry
that we will need. Particular emphasis will be on the ${\cal N}=2$ 
super Yang-Mills
theory and on susy non-linear $\s$-models describing effective theories.
The reader who is familiar with these matters 
may want to skip part or all of this 
section. In section 3, I review the analysis of Seiberg and Witten in its 
simplest setting: ${\cal N}=2$ susy YM theory with gauge group 
${\rm SU}(2)$ without additional ``matter" hypermultiplets. I will 
discuss the Wilsonian low-energy effective
action corresponding to this microscopic ${\cal N}=2$ super Yang-Mills action. 
The effective action describes the physics of the
remaining massless $U(1)$ susy multiplet in terms of an a priori {\it
unknown} function $\cf(a)$ where $a$ is the  vacuum expectation
value of the scalar field. 
${\cal N}=2$ supersymmetry constrains $\cf$ to be a (possibly multivalued) 
{\it holomorphic} function. Different vacuum expectation
values $a$ lead to physically different theories, and we have a moduli 
space with a complex coordinate $u$ that is related to $a$ again by a 
(possibly multivalued) holomorphic function $a(u)$.
Then I discuss how the function $\cf$ can be obtained in certain 
asymptotic regimes, using asymptotic freedom of the microscopic theory 
as well as duality of the effective low-energy theory. Technically, the 
asymptotic behaviours are translated into monodromy matrices describing how 
the couple $(a(u), \d \cf(a)/\d a)$ is transformed into itself as the 
coordinate $u$ goes once around the singular points of moduli space. 
Knowledge of the monodromy matrices and the asymptotics then allows to 
reconstruct the couple  $(a(u), \d \cf(a)/\d a)$ everywhere. This can 
be inverted,
at least in principle, to yield the function $\cf(a)$ and hence all knowledge 
about the low-energy effective action. However, this is not all we want to 
know. The low-energy effective action only describes the dynamics of the 
massless particles, namely the ``photon" supermultiplet. In addition there 
are the massive states, e.g. the analogues of the $W^\pm$ supermultiplets, the
magnetic monopoles, dyons,
etc. Even if we don't know their detailed dynamics, we can however determine 
their masses exactly at any point in moduli space, since they are BPS states 
and their masses are related to their charge quantum numbers and the functions 
$a(u)$ and $a_D(u)$. A more delicate question is to determine which BPS states 
are stable and exist in a given region of the moduli space. This requires the 
development of some simple new technique which was obtained in \cite{FB} and
will be explained in section 4 for the simplest example of the ${\rm SU}(2)$ 
gauge theory without hypermultiplets. Section 5 then generalises these results 
to the more complicated cases where various hypermultiplets are 
present \cite{BF}. 
In particular, if these hypermultiplets have bare masses one encounters a 
host of new phenomena \cite{BFmassive}, 
in particular the existence of Argyres-Douglas points 
where several mutually non-local fields simultaneously become massless and 
where the theory is superconformal.


\section{${\cal N}=2$ susy gauge theory}

We begin by giving a rather detailed review of those notions of 
supersymmetry that will be useful in the following. There are 
many reviews on supersymmetry, some of them are \cite{SUSY, gif, jpd}

\subsection{${\cal N}=1$ superspace}

A convenient and compact 
way to write actions for supersymmetric field theories is to introduce 
superspace and superfields, i.e. fields defined 
on superspace. This is particularly simple for unextended susy, so we will 
begin by looking at ${\cal N}=1$ superspace and superfields. Then we have two 
plus two 
susy generators $Q_\a$ and $\ov Q_\da$, as well as four generators $P_\m$ of 
space-time translations. There is one coordinate associated to each of them
so that coordinates on superspace are 
$(x^\m,\t_\a,\tb_\da)$ with $\t_\a$ and $tb_\da$, $\a, \da=1,2$ 
anticommuting as usual. 
We will use superspace as a very efficient tool to formulate supersymmetric 
theories.
We will not review the properties of superspace here but refer the reader 
to ref. \cite{gif} instead. Let us only remind the reader that a general 
superfield is
a function of all the coordinates on superspace, that the supercovariant 
derivatives are
defined as
\be\label{3covdiff}
\begin{array}{rcl}
D_\a &=&  {\d\over \d\t^\a} +i \s^\m_{\a\db} \tb^\db \d_\m \crbig
\Db_\da&=&=  {\d\over \d\tb^\da} + i \t^\b \s^\m_{\b\da} \d_\m  \cr
\end{array}
\ee
while the susy generators act as
\be\label{3susygendiff}
\begin{array}{rcl}
Q_\a &=& - i  {\d\over \d\t^\a} - \s^\m_{\a\db} \tb^\db \d_\m \crbig
\ov Q_\da&=& i {\d\over \d\tb^\da} + \t^\b \s^\m_{\b\da} \d_\m \ . \cr
\end{array}
\ee
They satisfy the susy algebra, in particular
\be\label{3susyanticom}
\{ Q_\a,  Q_\db\}= 2 \s^\m_{\a\db} P_\m = - 2i \s^\m_{\a\db} \d_\m
\ee
These generators then act on an arbitrary superfield $F$ as
\be\label{3susytransepsepsbar}
(1+i\e Q+i\eb \ov Q) F(x^\m,\t^\a,\tb^\db)
= F(x^\m -i\e\s^\m\tb+i\t\s^\m\eb, \t^\a+\e^\a, \tb^\db+\eb^\db)
\ee
and the susy variation of a superfield is of course defined as
\be\label{3susyvarF}
\delta_{\e,\eb} F= (i\e Q+i\eb \ov Q) F \ .
\ee

Since a general superfield contains too many component fields to 
correspond to an irreducible representation of ${\cal N}=1$ susy, it will 
be very useful to impose susy invariant condition to lower the number of
components. This can be done using the covariant derivatives $D_\a$ and 
$\Db_\da$ since they anticommute with the susy generators $Q$ and $\ov Q$. 
Then $\delta_{\e,\eb} (D_\a F) = D_\a (\delta_{\e,\eb} F)$ and idem for 
$\Db_\da$. It follows that $D_\a F=0$ or $\Db_\da F=0$ are susy 
invariant constraints one may impose to reduce the number of components
in a superfield.

\vskip 3.mm
\noindent
\underline{Chiral superfields}

\noindent
A chiral superfield $\f$ is defined by the condition
\be\label{3chiralsup}
\Db_\da \f = 0
\ee
and an anti-chiral one $\fb$ by $D_\a \fb = 0$. Introducing
$y^\m=x^\m+i\t\s^\m\tb \quad $ 
this is easily solved by 
\be\label{3chiralcompy}
\f(y,\t)=z(y)+\rd\t\p(y)-\t\t f(y)
\ee
or Taylor expanding  in terms of $x,\ \t$ and $\tb$:
\be\label{3chiralcompx}
\begin{array}{rcl}
\f(y,\t)&=&z(y)+\rd\t\p(y)-\t\t f(y) \crbig
&=&z(x) + \rd\t\p(x) + i\t\s^\m\tb \d_\m z(x) - \t\t f(x) 
-{i\over\rd} \t\t\d_\m\p(x) \s^\m\tb - {1\over 4}\t\t\tb\tb \d^2 z(x) \  \cr
\end{array}
\ee
and similarly for $\fb$.
Physically, such a chiral superfield describes one complex scalar $z$ and 
one Weyl fermion $\p$. The field $f$ will turn out to be an auxiliary field. 
The susy variations of the component fields are given by
\be\label{3susycompfieldstwo}
\begin{array}{rcl}
\delta z &=& \rd \e\p \crbig
\delta \p &=& \rd i \d_\m z \s^\m \eb  -\rd f \e \crbig
\delta f &=&  \rd i \d_\m \p  \s^\m \eb \ . \crbig
\end{array}
\ee

\vskip 3.mm
\noindent
\underline{Vector superfields}

\noindent
The ${\cal N}=1$ supermultiplet of next higher spin is the  vector multiplet. 
The corresponding superfield $V(x,\t,\tb)$ is real and has the expansion
\be\label{3vectorsuper}
\begin{array}{rcl}
V(x,\t,\tb)&=& C+i\t\chi-i\tb\cb +\t\s^\m\tb v_\m
+{i\over 2} \t\t (M+iN) -{i\over 2} \tb\tb (M-iN)\crbig
&+& i \t\t\, \tb\left( \lb +{i\over 2}\sb^\m\d_\m\chi\right)
- i \tb\tb\, \t\left( \l -{i\over 2}\s^\m\d_\m\cb\right) 
+\ha \t\t\tb\tb \left( D-\ha \d^2 C\right) \crbig
\end{array}
\ee
where all component fields only depend on $x^\m$. There are 8 bosonic 
components ($C,D,M,N,v_\m$) and 8 fermionic components ($\chi,\l$). 
These are too many components to describe a single supermultiplet. 
To reduce their number we make use of the supersymmetric generalisation 
of a gauge transformation. Note that the transformation
$V\to V+\f+\f^\dag$
with $\f$ a chiral superfield, implies the component transformation
$v_\m\to v_\m + \d_\m (2{\rm Im} z)$
which is an abelian gauge transformation. This shows that the 
transformation of $V$
indeed is the
desired supersymmetric generalisation of gauge transformations. 
If this transformation 
is a symmetry of the theory then, by an appropriate choice of $\f$, one can 
transform away the components $\chi,C,M,N$ and one component of $v_\m$. 
This choice is called the Wess-Zumino gauge, and it reduces the vector 
superfield to
\be\label{3vectWZ}
V_{\rm WZ}=\t\s^\m\tb v_\m(x) + i\t\t\, \tb\lb(x)
-i \tb\tb\, \t\l(x) +\ha \t\t\tb\tb D(x) \ .
\ee
Since each term contains at least one $\t$, the only non-vanishing power 
of $V_{\rm WZ}$ is
$V_{\rm WZ}^2=\t\s^\m\tb\ \t\s^\n\tb\ v_\m v_\n$  $= \ha \t\t\tb\tb\ v_\m v^\m$
and $V_{\rm WZ}^n=0$, $n\ge 3$.

To construct kinetic terms for the vector field $v_\m$ one must act on 
$V$ with the covariant derivatives $D$ and $\Db$. Define
\be\label{3walpha}
W_\a=-{1\over 4} \Db\Db D_\a V \quad , \quad 
\ov W_\da = -{1\over 4} DD \Db_\da V \ .
\ee
(This is appropriate for abelian gauge theories and will be slightly 
generalized in the non-abelian case.) Since $D^3=\Db^3=0$, $W_\a$ is 
chiral and $\ov W_\da$ antichiral. Furthermore it is clear that they 
behave as anticommuting Lorentz spinors. Note that they are invariant 
under the susy gauge transformation $V\to V+\f+\f^\dag$.
It is then easiest to use the WZ-gauge to 
compute $W_\a$. To facilitate things further, change variables to 
$y^\m,\t^\a,\tb^\da$. Then one finds
\be\label{3Walphacomp}
W_\a=-i \l_\a(y) + \t_\a D(y) + i (\s^{\m\n}\t)_\a f_{\m\n}(y) 
+ \t\t (\s^\m\d_\m \lb(y))_\a
\ee
with
\be\label{3abelianfmn}
f_{\m\n}=\d_\m v_\n - \d_\n v_\m
\ee
being the abelian field strength associated with $v_\m$.

\vskip 3.mm
\noindent
\underline{Susy invariant actions}

\noindent
To construct susy invariant actions we now only need to make a few 
observations. First, products of superfields are of course superfields. 
Also, products of (anti) chiral superfields are still (anti) chiral superfields.
Typically, one will have a superpotential $W(\f)$ which is again chiral. 
This $W$ may depend on several different $\f_i$. Using the $y$ and $\t$ 
variables one easily Taylor expands
\be\label{3superpot}
W(\f)= W(z(y)) +\rd {\d W\over \d z_i} \, \t\p_i(y) 
- \t\t \left( {\d W\over \d z_i} \, f_i(y)
+\ha {\d^2 W\over \d z_i  \d z_j} \, \p_i(y) \p_j(y) \right)
\ee
where it is understood that $\d W/\d z$ and $\d^2 W /\d z\d z$ are 
evaluated at $z(y)$.
The second and important observation is that any Lagrangian of the form
\be\label{3susylagr}
\Dint F(x,\t,\tb) + \Fint W(\f) + \Fbarint [ W(\f)]^\dag
\ee
is automatically susy invariant, i.e. it transforms at most by a total 
derivative in space-time. The proof is very simple and can be found 
e.g. in \cite{gif}.

As a first example consider an action for chiral superfields only
\be\label{3susyaction}
S= \int\, {\rm d}^4 x {\rm d}^2\t {\rm d}^2\tb \ \  \fb_i^\dag \f_i
+  \int\, {\rm d}^4 x {\rm d}^2\t \ W(\f_i) + h.c.
\ee
which in components  gives 
\be\label{3susyactiontwo}
S= \int\, {\rm d}^4 x \left[  
|\d_\m z_i|^2 -i \p_i\s^\m\d_\m \pb_i + f_i^\dag f_i 
- {\d W\over \d z_i} f_i + h.c. 
- \ha {\d^2 W\over \d z_i \d z_j} \p_i \p_j + h.c. \right] \ . 
\ee
More generally, one can replace $\f_i^\dag \f_i$ by a (real) K\"ahler 
potential $K(\f_i^\dag, \f_j)$. This leads to the non-linear $\s$-model 
discussed later. In any case, the $f_i$ have no kinetic term and hence 
are auxiliary fields. They should be eliminated by substituting their 
algebraic equations of motion
\be\label{3feoms}
f^\dag_i=\left( {\d W\over \d z_i}\right) 
\ee
into the action, leading to
\be\label{3susyactionthree}
S= \int\, {\rm d}^4 x \left[  
|\d_\m z_i|^2 -i \p_i\s^\m\d_\m \pb_i 
- \left|  {\d W\over \d z_i}\right|^2  
-  \ha {\d^2 W\over \d z_i \d z_j} \p_i \p_j 
- \ha \left( {\d^2 W\over \d z_i \d z_j}\right)^\dag \pb_i \pb_j 
\right] \ . 
\ee
We see that the scalar potential $V$ is determined in terms of the 
superpotential $W$ as
\be\label{3scalarpot}
V=\sum_i \left|  {\d W\over \d z_i}\right|^2 \ .
\ee

To construct an action for the vector superfield one remarks that,
since $W_\a$ is a chiral superfield, $\Fint W^\a W_\a$ will be 
a susy invariant Lagrangian. Its 
component expansion is obtained from the $\t\t$-term ($F$-term) of $W^\a W_\a$:
\be\label{3WWFterm}
W^\a W_\a \Big\vert_{\t\t}= -2i\l\s^\m\d_\m\lb + D^2 
- \ha (\s^{\m\n})^{\a\b} (\s^{\r\s})_{\a\b} f_{\m\n} f_{\r\s} \ ,
\ee
where we used $(\s^{\m\n})_\a^{\ \b}=\tr\,  \s^{\m\n}=0$. Furthermore,
$(\s^{\m\n})^{\a\b} (\s^{\r\s})_{\a\b}
=\ha\left( g^{\m\r}g^{\n\s}-g^{\m\s}g^{\n\r} \right) 
-{i\over 2} \e^{\m\n\r\s}$
(with $\e^{0123}=+1$) so that
\be\label{3gaugekinetic}
\Fint W^\a W_\a = -\ha f_{\m\n} f^{\m\n} 
-2i \l\s^\m\d_\m\lb + D^2 
+{i\over 4} \e^{\m\n\r\s} f_{\m\n} f_{\r\s}
\ .
\ee
Note that the first three terms are real while the last one is purely imaginary.

\subsection{${\cal N}=1$ susy YM action} 

We will now look at the non-abelian generalisation and construct the action
for ${\cal N}=1$ super YM theory coupled to matter multiplets 
which are chiral multiplets.
We  need a slight generalization of the definition of  
$W_\a$ to the non-abelian case.  All members of the vector multiplet (the 
gauge boson $v_\m$ and the gaugino $\l$) necessarily are in the same 
representation of the gauge group, i.e. in the adjoint representation. 
The chiral fields can be in any representation of the gauge group, e.g. in 
the fundamental one.
The non-abelian generalisation of the susy gauge transformation is
\be\label{4vecttrans}
e^{2gV} \to e^{i\L^\dag} e^{2gV} e^{-i\L} \ \Leftrightarrow
e^{-2gV} \to e^{i\L} e^{-2gV} e^{-i\L^\dag}
\ee
with $\L$ a chiral superfield and $g$ being the gauge coupling constant. 
This transformation can again be used 
to set $\chi,C,M,N$ 
and one component of $v_\m$ to zero, resulting in the same component 
expansion (\ref{3vectWZ}) of $V$ in the Wess-Zumino gauge. From now on we 
adopt this WZ gauge. Then  $V^n=0, n\ge 3$. The same remains true if some 
$D_\a$ or $\Db_\da$ are inserted in the product, e.g. $V (D_\a V) V=0$. 
One then simply has $e^{2gV}=1+2gV+ 2g^2 V^2$. The superfields $W_\a$ 
are now defined as
\be\label{4Wnonab}
W_\a=-{1\over 4} \Db \Db\left( e^{-2gV} D_\a e^{2gV}\right) \quad , \quad
\ov W_\da=+{1\over 4} D D \left( e^{2gV} \Db_\da e^{-2gV}\right) \ ,
\ee
which to first order in V reduces to the abelian definition. The
$W_\a$ now transform covariantly under the susy gauge transformations.
The component expansion of $W_\a$ in WZ gauge is given by 
\be\label{4Walphacomp}
{1\over 2g}\ W_\a=-i \l_\a(y) + \t_\a D(y) + i (\s^{\m\n}\t)_\a F_{\m\n}(y) 
+ \t\t (\s^\m D_\m \lb(y))_\a
\ee
where now
\be\label{3nonabelianfmn}
F_{\m\n}=\d_\m v_\n - \d_\n v_\m - ig [v_\m, v_\n]
\ee
and
\be\label{4covderiv}
D_\m \lb = \d_\m \lb  - ig  [v_\m, \lb] \ .
\ee
The reader should not confuse the gauge covariant derivative $D_\m$ neither 
with the super covariant derivatives $D_\a$ and $\Db_\da$, nor with 
the auxiliary field $D$.

The generators $T^a$ of the gauge group $G$ satisfy
\be\label{4structure}
[T^a,T^b]=i f^{abc} T^c
\ee
with real structure constants $f^{abc}$. The field strength then is 
$F^a_{\m\n}=\d_\m v^a_\n -\d_n v^a_\m +g f^{abc} v^b_\m v^c_\n$ and
the gauge covariant derivative is
$(D_\m\l)^a=\d_\m\l^a + g f^{abc}v^b_\m\l^c$.
One then introduces 
the complex coupling constant
\be\label{4tau}
\tau={\Theta\over 2\pi} +{4\pi i\over g^2}
\ee
where $\Theta$ stands for the $\Theta$-angle. (We use a capital $\Theta$ 
to avoid confusion with the superspace coordinates $\t$.) Then
\be\label{4gaugelagr}
\begin{array}{rcl}
{\cal L}_{\rm gauge} &=&
{1\over 32\pi} {\rm Im}\, \left( \tau \Fint {\rm Tr}\, W^\a W_\a \right)
\crbig
&=&{\rm Tr} \left( -{1\over 4} F_{\m\n}F^{\m\n} 
- i\l\s^\m D_\m \lb + \ha D^2\right)
+{\Theta\over 32\pi^2} g^2\, {\rm Tr} F_{\m\n} \widetilde F^{\m\n} \crbig
\end{array}
\ee
where
\be\label{4fdual}
\widetilde F^{\m\n} = \ha \e^{\m\n\r\s} F_{\r\s}
\ee
is the dual field strength. The single term $ {\rm Tr} W^\a W_\a $ has 
produced both, the conventionally normalised gauge kinetic term 
$-{1\over 4} {\rm Tr}F_{\m\n}F^{\m\n} $ and the instanton density 
${g^2\over 32\pi^2} {\rm Tr} F_{\m\n} \widetilde F^{\m\n}$ which multiplies 
the $\Theta$-angle!

We now add chiral (matter) multiplets $\f^i$ transforming in some 
representation $R$ of the gauge group where the generators are 
represented by matrices $(T^a_R)^i_{\ j}$. Then
\be\label{4mattertrans}
\f^i\to \left(e^{i\L}\right)^i_{\ j} \f^j \quad  , \quad
\f_i^\dag \to \f_j^\dag \left( e^{-i\L^\dag}\right)^j_{\ i}
\ee
or simply $\f\to e^{i\L}\f,\ \f^\dag\to \f^\dag e^{-i\L^\dag}$ where 
$\L=\L^a T^a_R $ is understood. Then 
\be\label{4ginvgen}
\f^\dag e^{2gV}\f\equiv \f^\dag e^{2g V^a T^a_R}\f \equiv 
\f_i^\dag \left( e^{2gV}\right)^i_{\ j} \f^j
\ee
is the gauge invariant 
generalisation of the kinetic term and
\be\label{4matterlagr}
{\cal L}_{\rm matter}=\Dint \f^\dag e^{2gV}\f +\Fint W(\f) 
+ \Fbarint [W(\f)]^\dag\ .
\ee
Working out the relevant superspace components yields
\be\label{4kinDtermrescaled}
\begin{array}{rcl}
\f^\dag e^{2g V} \f\Big\vert_{\t\t\tb\tb}&=& (D_\m z)^\dag D^\m z 
- i \p\s^\m D_\m \pb + f^\dag f \crbig
&+&i \rd  g z^\dag \l\p - i\rd  g \pb\lb z + g z^\dag D z
+ {\rm total \ derivative}\ . \crbig
\end{array}
\ee
now with $D_\m z=\d_\m z-i g v^a_\m T^a_R z$ and 
$D_\m\p=\d_\m\p- i g v^a_\m T^a_R\p$.
This part of the Lagrangian contains the kinetic terms for the scalar fields 
$z^i$ and the matter fermions $\p^i$, as well as specific interactions 
between the  $z^i$, the $\p^i$ and the gauginos 
$\l^a$. One has e.g. $z^\dag\l\p\equiv z_i^\dag (T^a_R)^i_{\ j} \l^a \p^j$.
What happens to the superpotential $W(\f)$? This must be a chiral superfield 
and hence must be constructed from the $\f^i$ alone. It must also be gauge 
invariant which imposes severe constraints on the superpotential. 
For the special case of ${\cal N}=2$ e.g. it will turn out that
no non-trivial superpotential is allowed. There is a last type of term that 
may appear in case the gauge group 
simply is ${\rm U}(1)$ or contains ${\rm U}(1)$ factors.
These are the 
Fayet-Iliopoulos terms. Since we will be mainly interested in groups without 
${\rm U}(1)$ factors we will not discuss them here.

We can finally write the full ${\cal N}=1$ Lagrangian, being the sum of 
(\ref{4gaugelagr}) and (\ref{4kinDtermrescaled}):
\be\label{4fullgaugelagr}
\begin{array}{rcl}
{\cal L}&=&  {\cal L}_{\rm gauge} + {\cal L}_{\rm matter}
\crbig
&=&
{1\over 32\pi} {\rm Im}\, \left( \tau \Fint {\rm Tr}\, W^\a W_\a \right)
+\Dint \f^\dag e^{2g V} \f
+\Fint W(\f) +\Fbarint [W(\f)]^\dag \crbig
&=&{\rm Tr} \left( -{1\over 4} F_{\m\n}F^{\m\n} 
- i\l\s^\m D_\m \lb + \ha D^2\right) 
+{\Theta\over 32\pi^2}g^2\, {\rm Tr} F_{\m\n} \widetilde F^{\m\n} 
\crbig
&+& (D_\m z)^\dag D^\m z 
- i \p\s^\m D_\m \pb + f^\dag f 
+i \rd  g z^\dag \l\p - i\rd  g \pb\lb z + g z^\dag D z \crbig
&-& {\d W\over \d z^i} f^i + h.c. 
- \ha {\d^2 W\over \d z^i \d z^j} \p^i \p^j + h.c. 
+ {\rm total \ derivative}\ . \crbig
\end{array}
\ee
The auxiliary field equations of motion are
\be\label{4fequ}
f_i^\dag= {\d W\over \d z^i}
\quad , \quad
D^a= - g z^\dag T^a z \ .
\ee
Substituting this back into the 
Lagrangian one finds
\be\label{4fullgaugelagrtwo}
\begin{array}{rcl}
{\cal L}&=& 
{\rm Tr} \left( -{1\over 4} F_{\m\n}F^{\m\n} 
- i\l\s^\m D_\m \lb\right) 
+{\Theta\over 32\pi^2}g^2\, {\rm Tr} F_{\m\n} \widetilde F^{\m\n}
+ (D_\m z)^\dag D^\m z 
- i \p\s^\m D_\m \pb  \crbig
&+&  i \rd  g z^\dag \l\p - i\rd  g \pb\lb z
- \ha {\d^2 W\over \d z^i \d z^j} \p^i \p^j 
- \ha \left({\d^2 W\over \d z^i \d z^j}\right)^\dag \pb^i \pb^j 
- V(z^\dag, z) + {\rm total \ derivative}\ , \crbig
\end{array}
\ee
where the scalar potential $V(z^\dag,z)$ is given by
\be\label{4scalarpot}
V(z^\dag, z)=f^\dag f + \ha D^2
=\sum_i \left| {\d W\over \d z^i}\right|^2
+{g^2\over 2} \sum_a \left|z^\dag T^a z\right|^2 \ .
\ee

\subsection{${\cal N}=1$ susy non-linear sigma model} 

As long as one wants to formulate a fundamental, i.e. microscopic theory, 
one is guided by the principle of renomalisability. For a gauge theory 
this is quite restrictive and the only possibility is the YM theory 
formulated above. The only freedom lies in the choice of gauge group and 
matter content, i.e. the number of chiral multiplets
and the representations 
of the gauge group under which they transform. Special choices will lead 
to extended supersymmetry, in particular ${\cal N}=2$ susy which will be 
discussed below. There is also some freedom to  choose the 
gauge invariant superpotential. Different choices will lead 
to different masses and Yukawa interactions.

In many cases, however, the theory one considers is an {\it effective} 
theory, valid at low energies only. Then renormalisability no longer is 
a criterion. The only restriction for such a low-energy effective theory  
is to contain no more than two (space-time) derivatives. Higher derivative 
terms are irrelevant at low energies. Thus we are led to study the 
supersymmetric non-linear sigma model. Another motivation comes from 
supergravity which is not renormalisable anyway. We will first consider 
the model for chiral multiplets only, and then extend the resulting theory 
to a gauge invariant one.

\vskip 3.mm
\noindent
\underline{Chiral multiplets only}

\noindent
We start with the action
\be\label{5sigmamod}
S=\xint \left( \Dint K(\f^i,\f_i^\dag) 
+\Fint w(\f^i) + \Fbarint w^\dag(\f_i^\dag)\right) \ .
\ee
We have denoted the superpotential by $w$ rather than $W$. The function 
$K(\f^i,\f_i^\dag)$ must be a real superfield, which will be the case if
$\ov K(z^i,z_j^\dag)= K(z_i^\dag,z^j)$. Derivatives with respect to its 
arguments will be denoted as
\be\label{5Kderiv}
K_i={\d\over \d z^i} K(z,z^\dag) \quad , \quad
K^j={\d\over \d z_j^\dag} K(z,z^\dag) \quad , \quad
K_i^j={\d^2\over \d z^i \d z_j^\dag} K(z,z^\dag)
\ee
etc. and
similarly $w_i={\d\over \d z^i} w(z) \quad , \quad
w_{ij}={\d^2\over \d z^i \d z^j} w(z)$
and $w^i=[w_i]^\dag$, $w^{ij}=[w_{ij}]^\dag$.

One has to expand the various terms in \eqn{5sigmamod}
and pick out the $\t\t\tb\tb$ terms 
or the $\t\t$ or $\tb\tb$ terms.  This is quite tedious and we refer to 
\cite{gif} for details. The result is
\be\label{5Dterms}
\Fint w(\f^i) + h.c. = \left( -w_i f^i -\ha w_{ij} \p^i \p^j \right) 
+ h.c.
\  \ee
and
\be\label{5KDtermtwo}
\begin{array}{rcl}
\Dint K(\f^i,\f_i^\dag)&=& 
K_i^j \left( f^i f_j^\dag + \d_\m z^i \d^\m z_j^\dag 
-{i\over 2} \p^i\s^\m\d_\m \pb_j +{i\over 2}\d_\m\p^i\s^\m\pb_j \right)
\crbig
&+&{i\over 4} K_{ij}^k \left( \p^i\s^\m\pb_k \d_\m z^j 
+ \p^j\s^\m \pb_k \d_\m z^i - 2i \p^i\p^j f_k^\dag \right) + h.c. 
\crbig
&+&{1\over 4} K_{ij}^{kl}  \p^i\p^j \, \pb_k\pb_l 
-{1\over 4} \d_\m\d^\m K(z^i,z_j^\dag)\ . 
\crbig
\end{array}
\ee
where the last term is a total derivative and hence can be dropped 
from the Lagrangian.
Note that after discarding this total derivative, (\ref{5KDtermtwo}) 
no longer contains the ``purely holomorphic" terms $\sim K_{ij}$ or 
the ``purely antiholomorphic" terms $\sim K^{ij}$. Only the mixed 
terms with at least one upper and one lower index
remain. This shows that the transformation
\be\label{5kahlertransf}
K(z,z^\dag) \to K(z, z^\dag) + g(z) + \ov g(z^\dag)
\ee
does not affect the Lagrangian. Moreover, the metric of the kinetic 
terms for the complex scalars is
\be\label{5kahlermetric}
K_i^j={\d^2\over \d z^i \d z^\dag_j} K(z, z^\dag)\ .
\ee
A metric like this obtained from a complex scalar function is called a K\"ahler
metric, and the scalar function $K(z, z^\dag)$ the K\"ahler potential. 
The metric is invariant under K\"ahler transformations (\ref{5kahlertransf}) 
of this potential. Thus one is led to interpret the complex scalars 
$z^i$ as (local) complex coordinates on a K\"ahler manifold, i.e. the 
target manifold of the sigma-model is K\"ahler. The K\"ahler invariance 
(\ref{5kahlertransf}) actually generalises to the superfield level
since
\be\label{5kahlertransfsuper}
K(\f, \f^\dag)\to K(\f, \f^\dag) + g(\f) + \ov g(\f^\dag) 
\ee
does not affect the resulting action because $g(\f)$ is again a chiral 
superfield and its $\t\t\tb\tb$ component is a total derivative, 
see (\ref{3chiralcompx}), hence $\Dint g(\f)=\Dint \ov g(\f^\dag)=0$.

Once $K_i^j$ is interpreted as a metric it is straightforward to 
compute the affine connection and curvature tensor. 
They are given by
\be\label{5gr}
\begin{array}{rcl}
\G_{ij}^l&=& (K^{-1})^l_k K^k_{ij} \quad , \quad 
\G^{ij}_l= (K^{-1})_l^k K_k^{ij} \ ,\crbig
R_{ij}^{kl}&=& K_{ij}^{kl} - K_{ij}^m (K^{-1})_m^n K_n^{kl} \ . \crbig
\end{array}
\ee
This allows us to rewrite various terms in the Lagrangian in a simpler and more
geometric form.
Define ``K\"ahler covariant" derivatives of the fermions as
\be\label{5fermcov}
\begin{array}{rcl}
D_\m \p^i &=& \d_\m \p^i + \G^i_{jk} \d_\m z^j\, \p^k 
=\d_\m \p^i + (K^{-1})^i_l\,  K^l_{jk} \d_\m z^j\, \p^k 
\crbig
D_\m \pb_j&=& \d_\m \pb_j + \G^{ki}_j \d_\m z^\dag_k\,  \pb_i
=\d_\m \pb_j + (K^{-1})_j^l\,  K^{ki}_l \d_\m z^\dag_k\,  \pb_i \ .
\crbig
\end{array}
\ee
The fermion bilinears in (\ref{5KDtermtwo}) then precisely are ${i\over 2}
K_i^j D_\m \p^i \s^\m \pb_j + h.c.$. The four fermion term is $K_{ij}^{kl}
\p^i\p^j\pb_k\pb_l$. The full curvature tensor will appear after we
eliminate the auxiliary fields $f^i$. To do this, we add the two pieces
(\ref{5KDtermtwo}) and  (\ref{5Dterms}) of the Lagrangian to see that the
auxiliary field equations of motion are 
\be\label{5auxequs}
f^i=(K^{-1})^i_j w^j - \ha \G_{jk}^i \p^j\p^k \ .
\ee
Substituting back into the sum of  (\ref{5KDtermtwo}) and  (\ref{5Dterms})  
we finally get the
Lagrangian
\be\label{5fulllagr}
\begin{array}{rcl}
& & \int {\rm d}^4 x \left[ \Dint K(\f, \f^\dag) 
+ \Fint w(\f) + \Fbarint [w(\f)]^\dag\right]
\crbig
& & = \int {\rm d}^4 x \Big[ 
K_i^j \left( \d_\m z^i \d^\m z^\dag_j + {i\over 2} D_\m \p^i \s^\m \pb_j 
- {i\over 2} \p^i \s^\m D_\m \pb_j \right) 
- (K^{-1})^i_j w_i w^j 
\crbig
& & \quad - \ha \left( w_{ij}- \G_{ij}^k w_k \right) \p^i\p^j
- \ha \left( w^{ij} - \G_k^{ij} w^k \right) \pb_i\pb_j 
+{1\over 4} R_{ij}^{kl} \p^i\p^j\pb_k\pb_l \Big] \ .
\crbig
\end{array}
\ee

\vskip 3.mm
\noindent
\underline{Including gauge fields}

\noindent
The inclusion of gauge fields changes two things. First, the kinetic term
$K(\f,\f^\dag)$ has to be modified so that, among others, all derivatives
$\d_\m$ are turned into gauge covariant derivatives as we did in the previous 
subsection
when we replaced $\f^\dag \f$ by $\f^\dag e^{2g V} \f$. Second, one has 
to add kinetic terms for the gauge multiplet $V$. In the spirit of the 
$\s$-model, one will allow a susy Lagrangian leading to terms of the form 
$f_{ab}(z) F^a_{\m\n} F^{b\m\n}$ etc.

Let's discuss the matter Lagrangian first. Since
\be\label{5gtran}
\f\to e^{i\L}\f \ , \quad
\f^\dag\to\f^\dag e^{-i\L^\dag} \ , \quad
e^{2gV}\to e^{i\L^\dag} e^{2gV} e^{-i\L}
\ee
one sees that
\be\label{5fegvtrans}
\f^\dag e^{2gV}\to \f^\dag e^{2gV} e^{-i\L} \ .
\ee
Then the combination $\left(\f^\dag e^{2gV}\right)_i \f^i$ is 
gauge invariant and the same is true for any real (globally) 
$G$-invariant function $K(\f^i, \f_i^\dag)$ if the argument 
$\f^\dag_i$ is replaced by 
$\left(\f^\dag e^{2gV}\right)_i$. We conclude that if $w(\f^i)$ 
is a $G$-invariant
function of the $\f^i$, i.e. if
\be\label{5wtrans}
w_i (T^a)^i_{\ j} \f^j =0 \quad , \quad a=1, \ldots {\rm dim}G
\ee
then
\be\label{5matterlagr}
{\cal L}_{\rm matter}=\Dint K\left( \f^i, \left(\f^\dag e^{2gV}\right)_i\right)
+\Fint w(\f^i) +\Fbarint [w(\f^i)]^\dag
\ee
is supersymmetric and gauge invariant.
To obtain the component expansion again is a bit lengthy. The result is
\be\label{5matterlagrcomp}
\begin{array}{rcl}
{\cal L}_{\rm matter}&=& K_i^j
\left[ f^i f_j^\dag + (D_\m z)^i (D^\m z)_j^\dag 
-{i\over 2} \p^i\s^\m \wt D_\m \pb_j +{i\over 2} \wt D_\m\p^i\s^\m\pb_j \right]
\crbig
&+&\ha K_{ij}^k\,  \p^i\p^j f_k^\dag + h.c. 
+{1\over 4} K_{ij}^{kl}\,  \p^i\p^j \, \pb_k\pb_l 
\crbig
&-&\left( w_i f^i +\ha w_{ij}\p^i\p^j\right) + h.c.
\crbig
&+& i\rd g K_j^i\, z_i^\dag \l \p^j - i\rd g K^i_j\, \pb_i \lb z^j 
+ g z_i^\dag D K^i \ .
\end{array}
\ee
Here all gauge indices have been suppressed, e.g. 
$\pb_i \lb z^j\equiv\pb_i T^a_R z^j \lb^a
\equiv (\pb_i)_M (T^a_R)^M_{\ N} (z^j)^N \lb^a$ where $(T^a_R)^M_{\ N}$ are 
the matrices of the
representation carried by the matter fields $(z^j)^N$ and $(\p^i)^N$. 
The derivatives $\wt D_\m$ acting
on the fermions are gauge and K\"ahler covariant, i.e.
\be\label{5fgaugekahlercov}
\begin{array}{rcl}
\wt D_\m \p^i &=& \d_\m \p^i -i g v^a_\m  T^a_R \p^i 
+ \G^i_{jk} \d_\m z^j\, \p^k  
\crbig
\wt D_\m \pb_j&=& \d_\m \pb_j -i g v^a_\m T^a_R \pb_j 
+ \G^{ki}_j \d_\m z^\dag_k\, 
\pb_i  \ .
\crbig
\end{array}
\ee

To discuss the generalisation of the gauge kinetic Lagrangian 
(\ref{4gaugelagr}),
reall that $W_\a$ is defined by (\ref{4Wnonab}) and in WZ 
gauge it
reduces to (\ref{4Walphacomp}). 
Note that any power of $W$ never contains more than
two derivatives, so we could consider a susy Lagrangian of the form 
$\Fint H(\f^i, W_\a)$ with an arbitrary $G$-invariant function $H$. 
We will be slightly less general and take
\be\label{5gaugelagr}
{\cal L}_{\rm gauge} = {1\over 16 g^2} \Fint f_{ab}(\f^i) W^{a\a}W^b_\a
+ h.c.
\ee
with $f_{ab}=f_{ba}$ transforming under $G$ as the symmetric product 
of the adjoint representation with itself. To get back the standard 
Lagrangian (\ref{4gaugelagr}) one only needs to take 
${1\over g^2} f_{ab} ={\tau\over 4\pi i} {\rm Tr}\, T^a T^b$,
so that $f_{ab}$ is identified with a matrix of generalised 
effective coupling constants.
Expanding (\ref{5gaugelagr}) in components is straightforward and yields
\be\label{5gaugelagrcomp}
\begin{array}{rcl}
{\cal L}_{\rm gauge}&=& {\rm Re} f_{ab}(z) 
\left( -{1\over 4} F^a_{\m\n}F^{b\m\n}
-i\l^a\s^\m D_\m \lb^b +\ha D^a D^b\right)
-{1\over 4} {\rm Im} f_{ab}(z) F^a_{\m\n} \wt F^{b\m\n} 
\crbig
&+&{1\over 4} f_{ab,i}(z) 
\left( \rd i \p^i\l^a D^b -\rd\l^a\s^{\m\n}\p^i F^b_{\m\n}
+\l^a\l^b f^i\right) +h.c.
\crbig
&+&{1\over 8} f_{ab,ij}(z) \l^a\l^b\p^i\p^j + h.c. 
\crbig
\end{array}
\ee
where $f_{ab,i}={\d\over \d z^i} f_{ab}(z)$ etc.

The full Lagrangian is given by 
${\cal L}={\cal L}_{\rm gauge}+{\cal L}_{\rm matter}$.
The auxiliary field equations of motion are
\be\label{5auxeoms}
\begin{array}{rcl}
f^i&=& (K^{-1})^i_j \left( w^j -\ha K^j_{kl}\, \p^k\p^l 
-{1\over 4} (f_{ab,j})^\dag \lb^a\lb^b \right) \crbig
D^a&=& - ({\rm Re} f)^{-1}_{ab} \left( g z^\dag_i T^b K^i 
+{i\over 2\rd} f_{bc,i}\p^i\l^c 
-{i\over 2\rd} (f_{bc,i})^\dag \pb_i \lb^c \right) \ .
\crbig
\end{array}
\ee
It is straightforward to substitute this into the Lagrangian 
${\cal L}$ and we will not write the result explicitly. Let us only 
note that the scalar potential is given by
\be\label{5sclarpot}
V(z,z^\dag)=(K^{-1})^i_j w_i w^j 
+{g^2\over 2}  ({\rm Re} f)^{-1}_{ab} (z^\dag_i T^a K^i) (z^\dag_j T^b K^j) \ .
\ee

\subsection{${\cal N}=2$ susy gauge theories} 

The ${\cal N}=2$ multiplets with helicities not exceeding one are the massless 
${\cal N}=2$ vector multiplet and the hypermultiplet. The former contains an 
${\cal N}=1$ vector multiplet and an ${\cal N}=1$ chiral multiplet, 
alltogether a 
gauge boson, two Weyl fermions and a complex scalar, while the 
hypermultiplet contains two ${\cal N}=1$ chiral multiplets. The 
${\cal N}=2$ vector
multiplet is necessarily massless while the hypermultiplet can be 
massless or be a short (BPS) massive multiplet. In this section 
we will concentrate 
on the ${\cal N}=2$ vector multiplet.
The ${\cal N}=2$ susy algebra is
\ba\label{n2alg}
\{Q^I_\a, \ov Q^J_\db \}&=&2 \s^\m_{\a\db} P_\m \delta^{IJ}\ , \crbig
\{Q^I_\a, Q^J_\b\}&=& 2 \sqrt{2} \e_{\a\b} \e^{IJ} Z\ , \crbig
\{\ov Q^I_\da, \ov Q^J_\db\} &=& 2 \sqrt{2} \e_{\da\db} \e^{IJ} Z^* \ .
\ea
In order to construct susy multiplets one combines these susy 
generators into fermionic harmonic oscillator operators. Positivity 
of the corresponding Hilbert space then requires
\be\label{massbound}
m\ge \sqrt{2} \vert Z \vert \ .
\ee
If the bound is satisfied, there are 2 combinations out of the 4 susy 
generators that
yield zero norm states. Hence these combinations should be set to zero, 
and effectively we are left with only half the susy generators. Thus if 
the bound is satisfied we have short multiplets with only four helicity 
states, while otherwise we have long multiplets with 16 helicity states. 
A short multiplet is called a BPS multiplet. For such a BPS state, 
the relation $m=\sqrt{2} \vert Z \vert$ between the 
mass and the central charge is very powerful \cite{WO}, since it allows 
for an exact determination of the mass, once 
its central charge is known.

\vskip 3.mm
\noindent
\underline{${\cal N}=2$ super Yang-Mills}

\noindent
Given the decomposition of the ${\cal N}=2$ vector multiplet into ${\cal N}=1$
 multiplets,
we start with a Lagrangian being the sum of the ${\cal N}=1$ gauge and matter 
Lagrangians (\ref{4gaugelagr}) and (\ref{4kinDtermrescaled}). At present, 
however, all fields are in the same ${\cal N}=2$ multiplet and hence must be in 
the same representation of the gauge group, namely the adjoint 
representation. The ${\cal N}=1$ matter Lagrangian (\ref{4kinDtermrescaled})
then becomes
\be\label{6matterlagr}
\begin{array}{rcl}
{\cal L}_{\rm matter}^{N=1}= \Dint \Tr \f^\dag e^{2gV} \f
&=& \Tr \Big[
(D_\m z)^\dag D^\m z 
- i \p\s^\m D_\m \pb + f^\dag f \crbig
&+&i \rd  g z^\dag \{\l,\p\} - i\rd  g \{\pb,\lb\} z + g D [z,z^\dag]
\Big]\, \crbig
\end{array}
\ee
where now
\be\label{6adjoint}
z=z^a T^a\ , \quad \p=\p^a T^a \ , \quad f= f^a T^a \ , 
\quad a=1, \ldots {\rm dim}G
\ee
in addition to $\l=\l^a T^a,\ D=D^a T^a,\ v_\m = v_\m^a T^a$. The commutators
or anticommutators arise since the generators in the adjoint representation 
are given by
\be\label{6adgen}
\left( T^a_{\rm ad}\right)_{bc} = -i f_{abc}
\ee
and we normalise the generators by
\be\label{6gennorm}
\Tr T^a T^b = \delta^{ab}
\ee
so that
\be\label{6trilin}
\begin{array}{rcl}
z^\dag \l\p &\to & z^\dag_b \l^a \left( T^a_{\rm ad}\right)_{bc}  \p^c
= -i  z^\dag_b \l^a f_{abc}\p^c =  i  z^\dag_bf_{bac}  \l^a \p^c \crbig
&=& z^\dag_b \l^a \p^c \Tr T^b [ T^a, T^c] = \Tr z^\dag \{ \l,\p\}
\end{array}
\ee
and
\be\label{6Dcomm}
z^\dag D z \to  z^\dag_b D^a \left( T^a_{\rm ad}\right)_{bc}  z^c
= -i  f_{abc} z^\dag_b D^a z^c
= - \Tr D[z^\dag, z] = \Tr D[z, z^\dag] \ .
\ee
We now add (\ref{6matterlagr}) to the ${\cal N}=1$ gauge lagrangian 
${\cal L}_{\rm gauge}^{N=1}$ (\ref{4gaugelagr})
and obtain
\be\label{6N=2lagr}
\begin{array}{rcl}
{\cal L}_{\rm YM}^{N=2} 
&=&{1\over 32\pi} {\rm Im}\, \left( \tau \Fint {\rm Tr}\, W^\a W_\a \right)
+ \Dint \Tr \f^\dag e^{2gV} \f \crbig
&=& \Tr  \Big( -{1\over 4} F_{\m\n}F^{\m\n} 
- i\l\s^\m D_\m \lb - i \p\s^\m D_\m \pb + (D_\m z)^\dag D^\m z \crbig
&&+
{\Theta\over 32\pi^2} g^2\, {\rm Tr} F_{\m\n} \widetilde F^{\m\n}
+ \ha D^2 + f^\dag f \crbig
&&+
i \rd  g z^\dag \{\l,\p\} - i\rd  g \{\pb,\lb\} z + g D [z, z^\dag]\Big) \ .
\crbig
\end{array}
\ee
A necessary and sufficient condition for ${\cal N}=2$ susy is the existence of an 
${\rm SU}(2)_R$ symmetry that rotates the two supersymmetry generators 
$Q^1_\a$ and $Q^2_\a$ 
into each other. As follows from the construction of the supermultiplet in 
section 2, the same symmetry must act between the two fermionic fields $\l$ 
and $\p$. Now the relative coefficients of ${\cal L}_{\rm gauge}^{N=1}$ and 
${\cal L}_{\rm matter}^{N=1}$ in (\ref{6N=2lagr}) have been chosen precisely 
in such a way to have this ${\rm SU}(2)_R$ symmetry: the $\l$ and $\p$ 
kinetic terms have the same coefficient, and the Yukawa couplings 
$z^\dag \{\l,\p\}$ and $\{\pb,\lb\} z$ also exhibit this symmetry. The 
Lagrangian (\ref{6N=2lagr}) is indeed ${\cal N}=2$ supersymmetric.

Note that we have not added a superpotential. Such a term (unless 
linear in $\f$) would  break the ${\rm SU}(2)_R$ invariance and not lead to 
an ${\cal N}=2$ theory.

The auxiliary field equations of motion are simply
\be\label{6auxeoms}
f^a=0 \quad , \quad
D^a= - g\, [z, z^\dag]^a
\ee
leading to a scalar potential
\be\label{6N=2scalarpot}
V(z,z^\dag)= \ha g^2\,  \Tr \left( [z,z^\dag]\right)^2 \ .
\ee
This scalar potential is fixed and a consequence solely of the auxiliary 
$D$-field of the ${\cal N}=1$ gauge multiplet.

\vskip 3.mm
\noindent
\underline{Effective ${\cal N}=2$ gauge theories}
 
\noindent
The above ${\cal N}=2$ super YM theory is renormalisable and constitutes the
asymptotically free microscopic theory we want to study below. However, we 
will be even more intersted in studying the effective low-energy action for 
the light degrees of freedom. As discussed above for the non-linear $\s$-model, if one considers effective theories, 
disregarding renormalisability, one may allow more general gauge and matter 
kinetic terms and start with an appropriate sum of (\ref{5matterlagr}) 
(with $w(\f^i)=0$) and (\ref{5gaugelagr}). It is clear however that the 
functions $f_{ab}$ cannot be independent from the K\"ahler potential $K$. 
Indeed, the ${\rm SU}(2)_R$ symmetry equates ${\rm Re} f_{ab}$ with the 
K\"ahler metric $K_a^b$. It turns out that this requires the following 
identification
\be\label{6prepot}
\begin{array}{rcl}
{16\pi\over (2g)^2}\, f_{ab}(z)
&=&-i {\d^2\over \d z^a \d z^b} \F(z) \equiv -i \F_{ab}(z) \crbig
{16\pi\over (2g)^2}\, K(z, z^\dag)
&=& - {i\over 2}\, z^\dag_a {\d\over \d z^a} \F(z) + h.c. \equiv 
- {i\over 2}\, z^\dag_a \F_a(z) + {i\over 2}\, \left[ \F_a(z)\right]^\dag z_a 
\crbig
\end{array}
\ee
where the holomorphic function $\F(z)$ is called the ${\cal N}=2$ prepotential. We 
have pulled out a factor ${16\pi\over (2g)^2} $ for later convenience. Also, 
we again absorb the factor $2g$ into the normalisation of the field. This 
makes sense since ${\rm Im} \F_{ab}$ will play the role of an effective 
generalised coupling. Hence we set 
\be\label{62g=1}
2g=1.
\ee
Then the full general ${\cal N}=2$ Lagrangian is
\be\label{6effN=2lagr}
\begin{array}{rcl}
{\cal L}_{\rm eff}^{N=2} 
&=& \left[ {1\over 64 \pi i} \Fint \F_{ab}(\f) W^{a\a} W^b_\a
+{1\over 32 \pi i}  \Dint \left( \f^\dag e^V \right)^a \F_a(\f) \right] + h.c. 
\crbig
&=& {1\over 16\pi} {\rm Im} \left[ \ha  \Fint \F_{ab}(\f) W^{a\a} W^b_\a + 
\Dint \left( \f^\dag e^V \right)^a \F_a(\f) \right] \ . \crbig
\end{array}\ee
Note that with the K\"ahler potential $K$ given by (\ref{6prepot}), 
the K\"ahler metric is proportional to ${\rm Im} \F_{ab}$ as required 
by ${\rm SU}(2)_R$ :
\be\label{6kahlermetric}
K_a^b={1\over 16\pi} {\rm Im} \F_{ab} 
={1\over 32\pi i} \left( \F_{ab}-\F_{ab}^\dag \right) \ .
\ee
The component expansion follows from the results of the previous section on 
the non-linear $\s$-model, using the identifications (\ref{6prepot}) and 
(\ref{6kahlermetric}), and taking vanishing superpotential $w(\f)$. In 
particular, the scalar potential is given by (cf (\ref{5sclarpot}))
\be\label{6effscalarpot}
V(z,z^\dag)= -{1\over 2\pi} ({\rm Im} \F)^{-1}_{ab} \, 
[z^\dag, \F_c(z)T^c]^a\, [z^\dag, \F_d(z) T^d]^b \ .
\ee

Let us insist that the full effective ${\cal N}=2$ action written in 
(\ref{6effN=2lagr}) is determined by a single holomorphic function 
$\F(z)$. Holomorphicity will turn out to be a very strong requirement. 
Finally note that $\F(z)= \ha \tau\, \Tr z^2$ 
gives back the  standard Yang-Mills Lagrangian (\ref{6N=2lagr}).

\section{Seiberg-Witten duality in ${\cal N}=2$ susy ${\rm SU}(2)$ 
gauge theory}

\def\a{\alpha}
\def\adot{{\dot\alpha}}
\def\b{\beta}
\def\m{\mu}
\def\s{\sigma}
\def\n{\nu}
\def\r{\rho}
\def\l{\lambda}
\def\lb{{\bar\lambda}}
\def\L{\Lambda}
\def\g{\gamma}
\def\G{\Gamma}
\def\t{\theta}
\def\tb{{\bar\theta}}
\def\sm{\sigma^\m}
\def\smb{\bar\sigma^\m}
\def\d{\partial}
\def\rmd{{\rm d}}
\def\vf{\varphi}
\def\f{\phi}
\def\F{\phi}
\def\fd{\phi_D}
\def\Fd{\phi_D}
\def\ix{\int {\rm d}^4 x \, }
\def\dt{{\rm d}^2 \t \, }
\def\dtb{{\rm d}^2 \tb \, }
\def\dtt{{\rm d}^2 {\tilde\t} \, }
\def\P{\Psi}
\def\cf{{\cal F}}
\def\cfd{{\cal F}_D}
\def\ad{a_D}
\def\e{\epsilon}
\def\rmd{{\rm d}}
\def\rd{\sqrt{2}}
\def\la{\langle}
\def\ra{\rangle}
\def\vac{\vert 0\rangle}
\def\P{\Psi}
\def\p{\psi}
\def\pb{{\bar \psi}}
\def\dd #1 #2{{\delta #1\over \delta #2}}
\def\tr{{\rm tr}\ }
\def\til{\widetilde}
\def\nm{\nabla_\m}
\def\fmn{F_{\m\n}}
\def\fmnt{{\tilde F}_{\m\n}}
\def\im{{\rm Im}\, }
\def\M{{\cal M}}

\subsection{Low-energy effective action of ${\cal N}=2$  $SU(2)$ YM theory}

Following Seiberg and Witten \cite{SW} 
we want to study and determine the low-energy
effective action of the ${\cal N}=2$ susy Yang-Mills theory with gauge group 
${\rm SU}(2)$. The latter theory is the microscopic theory which controls 
the high-energy behaviour. 
It was discussed in the previous subsection 
and its Lagrangian is given by (\ref{6N=2lagr}).
This theory is renormalisable and well-known to be asymptotically free. 
The low-energy effective
action will turn out to be quite different. Generalisations to 
bigger gauge groups have been extensively discussed in the 
literature, see e.g. \cite{LER, OGG}, but here we will restrict ourselves 
to the simplest case of ${\rm SU}(2)$. For other reviews, 
see e.g. \cite{ALG}.

\vskip 3.mm
\noindent
\underline{Low-energy effective actions}

\noindent
There are two types of effective actions. One is the standard generating 
functional
$\Gamma[\vf]$ of one-particle irreducible Feynman diagrams 
(vertex functions). It is
obtained from the standard renormalised generating functional $W[\vf]$ 
of connected
diagrams by a Legendre transformation. Momentum integrations in 
loop-diagrams are from
zero up to a UV-cutoff which is taken to infinity after renormalisation.
$\Gamma[\vf]\equiv \Gamma[\m,\vf]$ also depends on
the scale $\m$ used to define the renormalized vertex functions.

A quite different object is the Wilsonian effective action $S_{\rm W}[\m,\vf]$. It is
defined as $\Gamma[\m,\vf]$, except that all loop-momenta are only 
integrated down to $\m$ which serves as an infra-red cutoff. In theories with
massive particles only, there is no big difference between 
$S_{\rm W}[\m,\vf]$ and
$\Gamma[\m,\vf]$ (as long as $\m$ is less than the smallest mass). 
When massless particles are present, as is the case for gauge
theories, the situation is  different. In particular, in supersymmetric gauge
theories there is the so-called Konishi anomaly which can be viewed as 
an IR-effect. Although $S_{\rm W}[\m,\vf]$ depends holomorphically 
on $\m$, this is not the case for
$\Gamma[\m,\vf]$ due to this anomaly.

\vskip 3.mm
\noindent
\underline{The ${\rm SU}(2)$ case, moduli space}

\noindent
We want to determine the Wilsonian effective action in the case 
where the microscopic theory 
is the ${\rm SU}(2)$, ${\cal N}=2$ super Yang-Mills theory. As
explained above, classically this theory has a scalar potential 
$V(z)=\half g^2 \tr
([z^\dag,z])^2$ as given in (\ref{6N=2scalarpot}). 
Unbroken susy requires that $V(z)=0$ in the vacuum, but this still
leaves the possibilities of a vacuum with non-vanishing $z$
provided $[z^\dag,z]=0$.
We are interested in determining the gauge inequivalent vacua. 
A general $z$ is of the form 
$z(x)=\half\sum_{j=1}^3  \left( a_j(x)+i b_j(x)\right)
\sigma_j$ with real fields $a_j(x)$ and $b_j(x)$ 
(where I assume that not all three $a_j$
vanish, otherwise exchange the roles of the $a_j$'s and 
$b_j$'s in the sequel). By a
${\rm SU}(2)$ gauge transformation one can always arrange 
$a_1(x)=a_2(x)=0$.
Then $[z, z^\dag]=0$ implies
$b_1(x)=b_2(x)=0$ and hence, with $a= a_3 + i b_3$, one has 
$z=\half a \sigma_3$.
Obviously, in the vacuum $a$ must be a constant.
Gauge transformation from the Weyl group (i.e. rotations by
$\pi$ around the 1- or 2-axis of ${\rm SU}(2)$) can still change 
$a\to -a$, so $a$ and $-a$ are gauge
equivalent, too. The gauge invariant quantity describing inequivalent 
vacua is $\half
a^2$, or $\tr z^2$, which
 is the same, semiclassically.
When quantum fluctuations are important this is no longer so.
In the sequel, we will use the following definitions for $a$ and $u$:
\be\label{swmoduli}
u=\la \tr z^2\ra\quad , \quad \la z \ra = \half a\sigma_3\ .
\ee
The complex parameter $u$ labels gauge inequivalent vacua. The manifold 
of gauge
inequivalent vacua is called the moduli space $\M$ of the theory. Hence 
$u$ is a
coordinate on $\M$, and $\M$ is essentially the complex $u$-plane. We 
will see in the
sequel that $\M$ has certain singularities, and the knowledge of the 
behaviour of the
theory near the singularities will eventually allow the determination 
of the effective
action $S_{\rm W}$.

Clearly, for non-vanishing $\la z \ra$, the ${\rm SU}(2)$ gauge 
symmetry is broken by the
Higgs mechanism, since the $z$-kinetic term 
$\vert D_\m z\vert^2$ generates masses for
the gauge fields. With the above conventions, $v_\m^b,\ b=1,2$ become 
massive with masses given by $\half m^2 = g^2 \vert a\vert^2$, i.e 
$m=\rd  g \vert a\vert$. 
Similarly due to the $z,\l,\p$ interaction terms, $\p^b, \l^b,\ b=1,2$ 
become massive with the
same mass as the $v_\m^b$, as required by supersymmetry. Obviously, 
$v_\m^3,\p^3$ and
$\l^3$, as well as the mode of $z$ describing the flucuation of $z$ in the
$\sigma_3$-direction, remain massless. These massless modes are described by a
Wilsonian low-energy effective action which has to be ${\cal N}=2$ supersymmetry 
invariant,
since, although the gauge symmetry is broken, ${\rm SU}(2)\to {\rm U}(1)$, 
the ${\cal N}=2$ susy remains
unbroken. Thus it must be of the general form (\ref{6effN=2lagr})
where the indices $a,b$
now take only a single value ($a,b=3$) and will be suppressed since the gauge
group is ${\rm U}(1)$. Also,  in an
abelian theory there is no self-coupling of the gauge boson and the same 
arguments
extend to all members of the ${\cal N}=2$ susy multiplet: they do not carry 
electric charge.
Thus for a ${\rm U}(1)$-gauge theory, from (\ref{6effN=2lagr}) we simply get 
\be\label{swu1action}
{1\over 16\pi} \im\ix \left[ \ha \int\dt \cf''(\f)W^\a W_\a
+\int\dt\dtb \f^\dag\cf'(\f) \right] \ . 
\ee

\vskip 3.mm
\noindent
\underline{Metric on moduli space}

\noindent
As shown in (\ref{6kahlermetric}), the K\"ahler metric of the present 
$\s$-model is given by $K_{z \ov z} = {1\over 16\pi} \im
\cf''(z)$.
By the same token this defines the metric in the space of
(inequivalent) vacuum
configurations, i.e. the metric on moduli space as ($\bar a$ denotes the
complex conjugate of $a$)
\be\label{swmodspacemetric}
\rmd s^2=\im \cf''(a) \rmd a \rmd \bar a = \im \tau(a) \rmd a \rmd \bar a
\ee
where $\tau(a)=\cf''(a)$ is the effective (complexified) coupling constant 
according to the remark after eq. (\ref{5gaugelagr}). The $\s$-model 
metric $K_{z \ov z}$ has been replaced on
the moduli space $\M$ by ($16\pi$ times) its expectation value in the 
vacuum corresponding to the
given point on $\M$, i.e. by $\im \cf''(a)$.

The question now is whether the description of the effective action in 
terms of the
fields $\f, W$ and the function $\cf$ is appropriate for all vacua, i.e. 
for all value
of $u$, i.e. on all of moduli space. In particular the kinetic terms or
what is the same, the metric  on moduli space should be positive definite,
translating into $\im \tau(a) >0$. However, a simple argument shows that 
this cannot be
the case: since $\cf(a)$ is holomorphic, 
$\im\tau(a)=\im {\d^2\cf(a)\over \d a^2}$ is a
harmonic function and as such it cannot have a minimum, 
and hence (on the compactified
complex plane)  it cannot obey
$\im\tau(a)>0$ everywhere (unless it is a constant as in 
the classical case). The way
out is to allow for different local descriptions: the 
coordinates $a, \bar a$ and the
function $\cf(a)$ are appropriate  only in a certain region 
of $\M$. When a singular
point with $\im\tau(a)\to 0$ is approached one has to use a different set of
coordinates $\hat a$ in which $\im\hat\tau (\hat a)$ is non-singular (and
non-vanishing). This is possible provided the singularity of the metric is only a
coordinate singularity, i.e. the kinetic terms of the effective action are not
intrinsically singular, which will be the case.

\vskip 3.mm
\noindent
\underline{Asymptotic freedom and the one-loop formula}

\noindent
Classically the function $\cf(z)$ is given by $\half\tau_{\rm class} z^2$. 
The one-loop
contribution has been determined by Seiberg \cite{SEI}. 
The combined tree-level and one-loop result is
\be\label{tvi}
\cf_{\rm pert}(z)={i\over 2\pi} z^2\ln {z^2\over \L^2} \ .
\ee
Here $\L^2$ is some combination of $\m^2$ and numerical factors chosen 
so as to fix the
normalisation of $\cf_{\rm pert}$. Note that due to non-renormalisation 
theorems for
${\cal N}=2$ susy there are no corrections from two or more loops to the 
Wilsonian effective
action $S_{\rm W}$ and (\ref{tvi}) is the full perturbative result. 
There are however
non-perturbative corrections that will be determined below.

For very large $a$ the dominant contribution when computing 
$S_{\rm W}$ from the
microscopic ${\rm SU}(2)$ gauge theory comes from regions of large
 momenta ($p\sim a$) where the microscopic
theory is asymptotically free. Thus, as $a\to\infty$ the effective 
coupling constant goes to zero, and the perturbative expression 
(\ref{tvi}) for $\cf$ becomes an excellent
approximation. Also $u\sim \half a^2$ in this limit.\footnote{
One can check from the explicit solution below that one indeed has 
$\half a^2 - u = {\cal O}(1/u)$ as $u\to\infty$.
}
Thus
\be\label{tvii}
\begin{array}{rcl}
\cf(a  )&\sim& {i\over 2\pi} a^2\ln {a^2\over \L^2} \crbig
\tau(a) &\sim& {i\over \pi} \left( \ln {a^2\over \L^2} +3\right)  
\quad {\rm as}\
u\to\infty \ .\crbig
\end{array}
\ee
Note that due to the logarithm appearing at one-loop, 
$\tau(a)$ is a multi-valued
function of $a^2\sim 2u$. Its imaginary part, however, 
$\im\tau(a)\sim {1\over \pi}\ln{\vert a\vert^2\over \L^2}$ 
is single-valued and positive (for $a^2\to \infty$).

\subsection{Duality}

As already noted, $a$ and $\bar a$ do provide local coordinates on the 
moduli space
$\M$ for the region of large $u$. This means that in this region $\f$ and
$W^\a$ are appropriate fields to describe the low-energy effective action. 
As also
noted, this description cannot be valid globally, since $\im\cf''(a)$, 
being a harmonic
function, must vanish somewhere, unless it is a constant - which it is 
not. Duality will
provide a different set of (dual) fields $\f_{\rm D}$ and $W^\a_{\rm D}$ 
that provide an
appropriate description for a different region of the moduli space.

\vskip 3.mm
\noindent
\underline{Duality transformation}

\noindent
Define a dual field $\Fd$ and a dual function  $\cfd(\Fd)$ by
\be\label{qi}
\Fd=\cf'(\F) \quad , \quad
\cfd'(\Fd)=-\F \ .
\ee
These duality
transformations simply constitute a Legendre transformation
$\cfd(\Fd)$ $=\cf(\F)-\F\Fd$. Using these
relations, the second term in the $\f$ kinetic term of the action 
can be written as
\be\label{qiii}
\begin{array}{rcl}
\im\ix\dt\dtb\F^+\cf'(\F)&=&\im\ix\dt\dtb \left(-\cfd'(\Fd)\right)^+ \Fd \crbig
&=&
\im\ix\dt\dtb  \Fd^+ \cfd'(\Fd)\ .\crbig
\end{array}
\ee
We see that this second term in the effective action is invariant under the
duality transformation.

Next, consider the $\cf''(\F)W^\a W_\a$-term in the effective action 
(\ref{swu1action}). 
While the
duality transformation on $\F$ is local, this will not be the case for the
transformation of $W^\a$. Recall that $W$ contains the ${\rm U}(1)$ field 
strength $\fmn$.
This $\fmn$ is not arbitrary but of the form $\d_\m v_\n-\d_\n v_\m$ for
some $v_\m$. This can be translated into the Bianchi identity 
$\half\e^{\m\n\rho\sigma}
\d_\n F_{\rho\sigma}\equiv \d_\nu \tilde F^{\m\n}=0$. The 
corresponding constraint in
superspace is $\im (D_\a W^\a)=0$.  In the functional integral 
one has the choice of integrating
over $V$ only, or over $W^\a$ and imposing the constraint 
$\im (D_\a W^\a)=0$ by a real
Lagrange multiplier superfield which we call $V_D$:
\be\label{qiv}
\begin{array}{rcl}
&&\int{\cal D}V 
\exp\left[ {i\over 32\pi}\im\ix\dt\cf''(\F)W^\a W_\a\right]\crbig
&&\simeq \int{\cal D}W {\cal D}V_D
\exp\Big[ {i\over 32\pi}\im\ix\Big(\int\dt \cf''(\F)W^\a W_\a
+\ha \int\dt\dtb V_D D_\a W^\a\Big) \Big] \crbig
\end{array}
\ee
Observe that 
\be\label{qiva}
\begin{array}{rcl}
\int\dt\dtb V_D D_\a W^\a&=& - \int\dt\dtb D_\a V_D W^\a
=+\int\dt \bar D^2 (D_\a V_D W^\a)\crbig
&=&\int\dt (\bar D^2 D_\a V_D) W^\a
=-4\int\dt (W_D)_\a W^\a\crbig
\end{array}
\ee
where we used $\bar D_{\dot\b} W^\a=0$ and where the dual $W_D$ is 
defined from $V_D$
by $(W_D)_\a=-{1\over 4} \bar D^2 D_\a V_D$, as appropriate in the 
abelian case. 
Then one can do the
functional integral over $W$ and one obtains
\be\label{qv}
\int{\cal D}V_D \exp\left[ {i\over 32\pi}\im\ix\dt
\left( -{1\over\cf''(\F)} W_D^\a W_{D\a}\right)\right] \ .
\ee
This reexpresses the (${\cal N}=1$) supersymmetrized  Yang-Mills action in 
terms of a dual
Yang-Mills action with the effective coupling $\tau(a)=\cf''(a)$ 
replaced by $-{1\over\tau(a)}$. Recall that 
$\tau(a)={\theta(a)\over 2\pi}+{4\pi i \over g^2(a)}$, so that
$\tau\to -{1\over \tau}$ generalizes the inversion of the coupling 
constant discussed
in the introduction. Also, it can be shown that the replacement $W\to W_D$ 
corresponds to replacing
$\fmn\to \tilde F_{\m\n}$, the electromagnetic dual,  so that the manipulations
leading to (\ref{qv}) constitute a  duality transformation that 
generalizes the old electromagnetic duality of Montonen and Olive. 
Expressing the $-{1\over \cf''(\F)}$ in terms of $\Fd$ one sees from
(\ref{qi}) that $\cfd''(\Fd)=-{\rmd \F\over\rmd\Fd}=-{1\over\cf''(\F)}$ so that
\be\label{qvi}
-{1\over \tau(a)}=\tau_D(\ad) \ .
\ee
The whole action can then equivalently be written as
\be\label{qvii}
{1\over 16\pi} \im\ix \left[ \ha \int\dt \cfd''(\Fd)W_D^\a W_{D\a}
+\int\dt\dtb \Fd^+\cfd'(\Fd) \right] \ . 
\ee

\vskip 3.mm
\noindent
\underline{The duality group}

\noindent
To discuss the full group of duality transformations of the action it is most
convenient to write it as
\be\label{qviia}
{1\over 16\pi} \im \ix\dt {\rmd\Fd\over \rmd \F}W^\a W_\a +
{1\over 32 i \pi} \ix\dt\dtb \left( \F^+\Fd-\Fd^+\F\right) \ .
\ee
While we have shown in the previous subsection that there is a duality symmetry
\be\label{qviii}
\pmatrix{\Fd\cr \F\cr} \to \pmatrix{0&1\cr -1&0\cr} \pmatrix{\Fd\cr \F\cr} \ ,
\ee
the form (\ref{qviia}) shows that there  also is a symmetry
\be\label{qix}
\pmatrix{\Fd\cr \F\cr} \to \pmatrix{1&b\cr 0&1\cr} 
\pmatrix{\Fd\cr \F\cr}\quad ,
\quad b\in {\bf Z} \ .
\ee
Indeed,  in (\ref{qviia}) the second term remains invariant since
$b$ is real, while the first term 
gets shifted by
\be\label{qx}
{b\over 16\pi} \im\ix\dt W^\a W_\a = - {b\over 16\pi}  
\ix \fmn\tilde F^{\m\n} =-2\pi
b \n
\ee
where $\n\in {\bf Z}$ is the instanton number. Since the action appears 
as $e^{iS}$ in
the functional integral, two actions differing only by $2\pi  {\bf Z}$ 
are equivalent,
and we conclude that (\ref{qix}) with integer $b$ is a symmetry of 
the effective action. The
transformations (\ref{qviii}) and (\ref{qix}) together generate the 
group $Sl(2,{\bf Z})$. This is
the group of duality symmetries.

Note that the metric (\ref{swmodspacemetric}) on moduli space can be 
written as
\be\label{qxa}
\rmd s^2 =\im(\rmd \ad\rmd \bar a) 
= {i\over 2} (\rmd a\rmd\bar \ad - \rmd \ad \rmd
\bar a)
\ee
where $\ad=\d\cf(a)/\d a$, and that this metric
obviously also is invariant under the duality group  $Sl(2,{\bf Z})$

\vskip 3.mm
\noindent
\underline{Monopoles, dyons and the BPS mass spectrum}

\noindent
At this point, I will have to add a couple of ingredients without much further
justification and refer the reader to the literature for more details.

In a spontaneously broken gauge theory as the one we are considering, 
typically there
are solitons (static, finite-energy solutions of the equations of motion) 
that carry
magnetic charge and behave like non-singular magnetic monopoles (for a
pedagogical treatment, see Coleman's lectures \cite{COL}). 
The duality transformation 
(\ref{qviii}) constructed above
exchanges electric and magnetic degrees of freedom, hence electrically 
charged states,
as would be described by hypermultiplets of our ${\cal N}=2$ supersymmetric 
version, with magnetic monopoles.

As for any theory with extended supersymmetry, there are long and short (BPS)
multiplets in the present ${\cal N}=2$ theory.
small (or short) multiplets have 4 
helicity states and large (or long) ones have 16 helicity states. 
As discussed earlier, massless states must be 
in short
multiplets, while massive states are in short ones if they satisfy the 
BPS condition
$m^2=2\vert Z\vert^2$, or in long ones if
$m^2>2\vert Z\vert^2$. Here $Z$ is the central charge of the ${\cal N}=2$ 
susy algebra.
The states that become massive by the Higgs mechanism must
be in short multiplets since they were before the symmetry breaking 
and the Higgs mechanism cannot generate
the missing $16-4=12$ helicity states. The heavy gauge bosons\footnote{
Again, to conform with the Seiberg-Witten normalisation, we have 
absorbed a factor 
of $g$ into $a$ and $a_D$, so that the masses of the heavy gauge bosons 
now are $m=\rd |a|$  rather than $\rd g |a|$.}
have masses $m=\rd |a|= \rd |Z|$ and hence $Z=a$. This generalises to all 
purely electrically charged states as
$Z=a n_e$ where $n_e$ is the (integer) electric charge. Duality then 
implies that a
purely magnetically charged state has $Z=\ad (-n_m)$ 
where $n_m$ is the (integer)
magnetic charge. A state with both types of charge, called a dyon,
 has $Z=a n_e - \ad n_m$ since the
central charge is additive. All this applies to states in short multiplets, 
so-called BPS-states. The mass formula  for these states then is
\be\label{qxi}
m^2=2\vert Z\vert^2\quad , \quad 
Z= a n_e - \ad n_m 
= (n_e,n_m) \pmatrix{0&1\cr -1&0\cr} \pmatrix{\ad\cr a\cr} 
\equiv \eta\left( \pmatrix{n_e\cr n_m\cr}, \pmatrix{\ad\cr a\cr}\right)  \ 
\ee
where $\eta$ is the standard symplectic product such that for any
$Sl(2,{\bf Z})\equiv Sp(2,{\bf Z})$ transformation
$M=\pmatrix{\a&\b\cr\g&\delta\cr}$ acting on $\pmatrix{\ad\cr a\cr}$
one has
\be\label{sympleta}
\eta\left( \pmatrix{n_e\cr n_m\cr}, M\, \pmatrix{\ad\cr a\cr}\right)
=\eta\left( M^{-1}\, \pmatrix{n_e\cr n_m\cr}, \pmatrix{\ad\cr a\cr}\right)
\ .
\ee
It is then clear that under such an $Sl(2,{\bf Z})$ transformation $M$ the
charge vector gets
transformed to $ M^{-1}\, (n_m,n_e) =  (n'_m,n'_e)$ with 
integer $n'_e$ and $n'_m$.
In particular, one sees again at the level of the charges that
the transformation (\ref{qviii}) exchanges purely electrically charged 
states with purely
magnetically charged ones. It can be shown (section 4.1)
that precisely those BPS
states are stable for which $n_m$ and $n_e$ are  relatively prime, i.e. 
for stable states $(n_m,n_e) \ne (qm,qn)$ for integer $m,n$ and $q\ne \pm 1$.

\subsection{Singularities and Monodromy}

In this section we will study the behaviour of $a(u)$ and $\ad(u)$ as 
$u$ varies on the
moduli space $\M$. Particularly useful information will be obtained 
from their
behaviour as $u$ is taken around a closed contour. If the contour 
does not encircle
certain singular points to be determined below, $a(u)$ and $\ad(u)$ 
will return to
their initial values once $u$ has completed its contour. However, 
if the $u$-contour
goes around these singular points, $a(u)$ and $\ad(u)$ do not return 
to their initial
values but rather to certain linear combinations thereof: one has a non-trivial
monodromy for the multi-valued functions $a(u)$ and $\ad(u)$.

\vskip 3.mm
\noindent
\underline{The monodromy at infinity}

\noindent
This is immediately clear from the  behaviour near $u=\infty$. As already
explained in section 3.4, as $u\to \infty$, due to asymptotic freedom,
the perturbative expression for $\cf(a)$ is valid and one has from (\ref{tvi})
for $\ad=\d\cf(a)/\d a$
\be\label{ci}
\ad(u)={i\over \pi} a \left( \ln{a^2\over \L^2}+1\right)\quad , \quad
u\to\infty \ .
\ee
Now take $u$ around a counterclockwise contour of very large radius in
the complex $u$-plane, often simply written as $u\to e^{2\pi i}u$. This
is equivalent to having $u$ encircle the point at $\infty$ on the Riemann
sphere in a {\it clockwise} sense. In any case, since $u=\half a^2$ (for
$u\to\infty$) one has $a\to -a$ and
\be\label{cii}
\ad\to {i\over\pi} (-a) \left( \ln{e^{2\pi i}a^2\over \L^2}+1\right)
=-\ad+2a
\ee
or
\be\label{ciii}
\pmatrix{\ad(u)\cr a(u)\cr}\to M_\infty  \pmatrix{\ad(u)\cr a(u)\cr}
\quad , \quad M_\infty=\pmatrix{-1&2\cr 0&-1\cr} \ .
\ee
Clearly, $u=\infty$ is a branch point of $\ad(u)\sim {i\over
\pi}\sqrt{2u} \left(\ln{u\over\L^2}+1\right)$. This is why this point is 
referred to as a
singularity of the moduli space.

\vskip 3.mm
\noindent
\underline{How many singularities?}

\noindent
Can $u=\infty$ be the only singular point? Since a branch cut has to
start and end somewhere, there must be at least one other singular point.
Following Seiberg and Witten, I will argue that one actually needs three
singular points at least. To see why two cannot work, let's suppose for a
moment that there are only two singularities and show that this leads to
a contradiction.

Before doing so, let me note that there is an important 
so-called ${\rm U}(1)_R$-symmetry in the classical theory
that takes $z\to e^{2i\a}z$, $\f \to e^{2i\a}\f$,
$W \to e^{i\a}W$, $\t\to e^{i\a}\t$,
$\tb\to e^{i\a}\tb$, thus $\dt \to e^{-2i\a} \dt$,
$\dtb \to e^{-2i\a}\dtb$. Then the
classical action is invariant under this global
symmetry. More generallly, the action  will be invariant if
$\cf(z) \to e^{4i\a} \cf(z)$. This symmetry is broken by the one-loop
correction and also by instanton contributions. The latter give
corrections to $\cf$ of the form $z^2\sum_{k=1}^\infty c_k \left( \L^2/
z^2 \right)^{2k}$, and hence are invariant only for 
$\left(e^{4i\a}\right)^{2k} =1$, i.e. $\a={2\pi n\over 8},\ n\in{\bf Z}$.
Hence instantons break the ${\rm U}(1)_R$-symmetry  to a dicrete ${\bf Z}_8$.
The one-loop corrections behave as 
${i\over 2\pi}z^2\ln {z^2\over\L^2}\to e^{4i\a}\left(
{i\over 2\pi}z^2\ln {z^2\over\L^2} - {2\a\over \pi}z^2\right)$. As 
before one shows that this only changes the
action by $2\pi\n \left({4\a\over \pi}\right)$ where $\n$ is integer, so
that again this change is irrelevant as long as ${4\a\over \pi}=n$ or
$\a={2\pi n\over 8}$. Under this ${\bf Z}_8$-symmetry, $z\to e^{i\pi
n/2}z$, i.e. for odd $n$ one has $z^2\to -z^2$. The non-vanishing
expectation value $u=\la\tr z^2\ra$ breaks this ${\bf Z}_8$ further to
${\bf Z}_4$. Hence for a given vacuum, i.e. a given point on moduli space
there is only a ${\bf Z}_4$-symmetry left from the ${\rm U}(1)_R$-symmetry.
However, on the manifold of all possible vacua, i.e. on $\M$, one has
still the full ${\bf Z}_8$-symmetry, taking $u$ to $-u$. Said 
differently, the quotient ${\bf Z}_8/{\bf Z}_4 = {\bf Z}_2$ acts 
as a symmetry on moduli space mapping the theory at $u$ to the 
theory at $-u$.

Due to this global symmetry $u\to -u$, singularities of $\M$ should come
in pairs: for each singularity at $u=u_0$ there is another one at
$u=-u_0$. The only fixed points of $u\to -u$ are $u=\infty$ and $u=0$. We
have already seen that $u=\infty$ is a singular point of $\M$. So if
there are only two singularities the other must be the fixed point $u=0$.

If there are only two singularities, at $u=\infty$ and $u=0$, then by contour
deformation (``pulling the contour over the back of the sphere")\footnote{
It is well-known from complex analysis that monodromies are associated 
with contours
around branch points. The precise from of the contour does not matter, 
and it can be
deformed as long as it does not meet another branch point. Our singularities 
precisely
are the branch points of $a(u)$ or $\ad(u)$.}
 one sees that the
monodromy around 0 (in a counterclockwise sense) is the same as the above 
monodromy
around $\infty$: $M_0=M_\infty$. But then $a^2$ is not affected by any
monodromy and
hence is a good global coordinate, so
one can take $u=\half a^2$ on all of $\M$, and furthermore one must have
\be\label{civ}
\begin{array}{rcl}
\ad&=&{i\over \pi} a \left( \ln {a^2\over \L^2}+1\right) + g(a)\crbig
a&=&\sqrt{2u}\crbig
\end{array}
\ee
where $g(a)$ is some entire function of $a^2$. This implies that
\be\label{cv}
\tau={\rmd\ad\over\rmd a}={i\over \pi} \left( \ln {a^2\over \L^2}+3\right) +
{\rmd g\over\rmd a} \ .
\ee
The function $g$ being entire, $\im {\rmd g\over\rmd a}$ cannot have a 
minimum (unless
constant) and it is clear that $\im\tau$ cannot be positive everywhere. 
As already
emphasized, this means that $a$ (or rather $a^2$) cannot be a good global 
coordinate and
(\ref{civ}) cannot hold globally. Hence, two singularities only cannot work.

The next simplest choice is to try 3 singularities. Due to the $u\to -u$ 
symmetry,
these 3 singularities are at $\infty, u_0$ and $-u_0$ for some $u_0\ne 0$. 
In particular, $u=0$ is no
longer a singularity of the quantum moduli space. To get a singularity 
also at $u=0$
one would need at least four singularities at $\infty, u_0, -u_0$ and $0$. 
As discussed
later, this is not possible, and more generally, exactly 3 singularities 
seems to be
the only consistent possibility.

So there is no singularity at $u=0$ in the quantum moduli space $\M$.
Classically, however, one precisely expects that $u=0$ should be a 
singular point,
since classically $u=\half a^2$, hence $a=0$ at this point, and
then there is no Higgs mechanism any more. Thus all (elementary) 
massive states, i.e.
the gauge
bosons $v_\m^1, v_\m^2$ and their susy partners $\p^1, \p^2, \l^1, \l^2$
become massless. Thus the description of the lights fields in terms of
the previous Wilsonian effective action should break down, inducing a 
singularity on
the moduli space. As already stressed, this is the clasical picture. 
While $a\to
\infty$ leads to asymptotic freedom  and the microscopic ${\rm SU}(2)$ 
theory is weakly
coupled, as $a\to 0$ one goes to a strong coupling regime where the 
classical
reasoning has no validity any more, and $u\ne \half a^2$. By the BPS 
mass formula 
(\ref{qxi})
massless gauge bosons still are possible at $a=0$, but this does no 
longer correspond to
$u=0$.

So where has the singularity due to massless gauge bosons at $a=0$ 
moved to? One might be tempted to think that
$a=0$ now corresponds to the singularities at $u=\pm u_0$, but this is 
not the case
as I will show in a moment. The answer is that the point $a=0$ no
longer belongs to the quantum moduli space (at least not to the component 
connected to
$u=\infty$ which is the only thing one considers). This can be seen 
explicitly from
the form of the solution for $a(u)$ given in the next section.

\vskip 3.mm
\noindent
\underline{The strong coupling singularities}

\noindent
Let's now concentrate on the case of three singularities at 
$u=\infty, u_0$ and $-u_0$.
What is the interpretation of the (strong-coupling) singularities 
at finite $u=\pm
u_0$? One might first try to consider that they are still due to 
the gauge bosons
becoming massless. However, as Seiberg and Witten point out, 
massless gauge bosons
would imply an asymptotically conformally invariant theory in 
the infrared limit and
conformal invariance implies $u=\la\tr z^2\ra=0$ unless $\tr z^2$ 
has dimension zero
and hence would be the unity operator - which it is not. So the 
singularities at $u=\pm
u_0\ (\ne 0)$ do not correspond to massless gauge bosons.

There are no other elementary ${\cal N}=2$ multiplets in our theory. 
The next thing to try
is to consider collective excitations - solitons, like the 
magnetic monopoles or dyons.
Let's first study what happens if a magnetic monopole  of 
unit magnetic charge becomes
massless. From the BPS mass formula (\ref{qxi}), the mass of 
the magnetic monopole is
\be\label{cvi}m^2=2\vert \ad\vert^2
\ee
and hence vanishes at $\ad=0$. We will see that this produces one of the two
strong-coupling singularities. So call $u_0$ the value of $u$ at which
$\ad$ vanishes.
Magnetic monopoles are described by hypermultiplets $H$ of ${\cal N}=2$ susy 
that couple
locally to the dual fields $\Fd$ and $W_D$, just as electrically 
charged ``electrons"
would be described by hypermultiplets that couple locally to $\F$ 
and $W$. So in the
dual description we have $\Fd, W_D$ and $H$, and, near $u_0$, 
$\ad\sim \la \Fd\ra$ is
small. This theory is exactly ${\cal N}=2$ susy QED with very light 
electrons (and a subscript
$D$ on every quantity). The latter theory is not asymptotically 
free, but has a
$\b$-function given by
\be\label{cvii}
\m{\rmd\over \rmd\m} g_D={g_D^3\over 8\pi^2}
\ee
where $g_D$ is the coupling constant. But the scale $\m$ is 
proportional to $\ad$ and
${4\pi i\over g_D^2(\ad)}$ is $\tau_D$ for $\t_D=0$ (of course, 
super QED, unless
embedded into a larger gauge group,  does not
allow for a non-vanishing theta angle). One concludes that for 
$u\approx u_0$ or $\ad\approx 0$
\be\label{cviii}
\ad {\rmd\over \rmd \ad} \tau_D=-{i\over \pi} \ 
\Rightarrow \ \tau_D = -{i\over \pi}
\ln\ad \ .
\ee
Since $\tau_D={\rmd (-a)\over \rmd\ad}$ this can be integrated to give
\be\label{cix}
a\approx a_0+{i\over \pi} \ad\ln\ad \qquad (u\approx u_0)
\ee
where we dropped a subleading term $-{i\over \pi}\ad$. Now, $\ad$ 
should be a good
coordinate in the vicinity of $u_0$, hence depend linearly\footnote{
One might want to try a more general dependence like 
$\ad\approx c_0 (u-u_0)^k$ with
$k>0$. This leads to a monodromy in $Sl(2,{\bf Z})$ only for integer $k$. 
The factorisation
condition below, together with the form of $M(n_m,n_e)$ also given below, 
then imply that
$k=1$ is the only possibility.
}
on $u$. One concludes
\be\label{cx}
\ad\approx c_0 (u-u_0)\quad , \quad
a\approx  a_0+{i\over \pi} c_0 (u-u_0)\ln (u-u_0) \ .
\ee
From these expressions one immediately reads the monodromy as $u$ turns 
around $u_0$
counterclockwise, $u-u_0\to e^{2\pi i} (u-u_0)$:
\be\label{cxi}
\pmatrix{\ad\cr a\cr} \to \pmatrix{ \ad\cr a-2\ad \cr} = M_{u_0} 
\pmatrix{\ad\cr a\cr} \quad , \quad
 M_{u_0}=\pmatrix{1&0\cr -2&1\cr} \ . 
\ee
Note that the magnetic monopole 
$\pmatrix{n_e\cr n_m\cr} = \pmatrix{0\cr 1\cr}$ is invariant 
under this monodromy, i.e. it is an eigenvector of $M_{u_0}$ 
with unit eigenvalue.

To obtain the monodromy matrix at $u=-u_0$ it is enough to observe 
that the contour
around $u=\infty$ is equivalent to a counterclockwise contour of 
very large radius in
the complex plane. This contour can be deformed into a contour 
encircling $u_0$ and a
contour encircling $-u_0$, both counterclockwise. It follows the 
factorisation
condition on the monodromy matrices\footnote{
There is an ambiguity concerning the ordering of $M_{u_0}$ and $M_{-u_0}$
which will be resolved below.}
\be\label{cxii}
M_\infty=M_{u_0} M_{-u_0}
\ee
and hence
\be\label{cxiii}
M_{-u_0} = \pmatrix{-1&2\cr -2&3\cr} \ .
\ee

What is the interpretation of this singularity at $u=-u_0$? 
As discussed above,
using the  $Sl(2,{\bf Z})$ invariance of $Z$, the monodromy transformation
$\pmatrix{\ad\cr a\cr}\to M \pmatrix{\ad\cr a\cr}$ can be 
interpreted as changing the
magnetic and electric quantum numbers as
$\pmatrix{n_e\cr n_m\cr} \to M^{-1} \pmatrix{n_e\cr n_m\cr}$.
The state of vanishing mass responsible for a singularity 
should be invariant under the
monodromy, and hence be an eigenvector of $M$ with unit 
eigenvalue. We already noted this
for the magnetic monopole. Similarly, the eigenvector 
of (\ref{cxiii}) with unit eigenvalue
is $(n_e, n_m)=(1,1)$. This is a dyon. Thus the sigularity at 
$-u_0$ is interpreted as
being due to a $(1,1)$ dyon becoming massless.

More generally, $(n_e, n_m)$ is the eigenvector with unit 
eigenvalue\footnote{
Of course, the same is true for any $(q n_m, q n_e)$ with 
$q\in {\bf Z}$, but according
to the discussion in section 4.3 on the stability of BPS states, 
states with $q\ne \pm
1$ are not stable.}
of
\be\label{cxv}
M(n_e,n_m)=\pmatrix{ 1-2n_m n_e& 2 n_e^2\cr -2 n_m^2 & 1+ 2 n_m n_e\cr}
\ee
which is the monodromy matrix that should appear for any 
singularity due to a massless
dyon with charges $(n_m, n_e)$. Note that $M_\infty$ as 
given in (\ref{ciii}) is not of this
form, since it does not correspond to a hypermultiplet becoming massless.

One notices that the relation (\ref{cxii}) does not look 
invariant under $u\to -u$, i.e
$u_0\to -u_0$ since $M_{u_0}$ and $M_{-u_0}$ do not commute. 
The apparent contradiction
with the ${\bf Z}_2$-symmetry is resolved by the following remark. 
The precise
definition of the composition of two monodromies as in (\ref{cxii}) 
requires a choice of
base-point $u=P$ (just as in the definition of homotopy groups). 
Using a different
base-point, namely $u=-P$, leads to 
\be\label{cxvi}
M_\infty =M_{-u_0}M_{u_0}
\ee
instead. Then one would obtain $M_{-u_0}=\pmatrix{3&2\cr -2&-1}$, 
and comparing with
(\ref{cxv}), this would be interpreted as due to a $(-1,1)$ dyon. Thus the 
${\bf Z}_2$-symmetry $u\to -u$ on the quantum moduli space also 
acts on the base-point $P$,
hence exchanging (\ref{cxii}) and (\ref{cxvi}). At the same time 
it exchanges the $(1,1)$ dyon
with the $(-1,1)$ dyon.

Does this mean that the $(1,1)$ or  $(-1,1)$ dyons play a 
privileged role? Actually
not. If one first turns $k$ times around $\infty$, then around 
$u_0$, and then $k$
times around $\infty$ in the opposite sense, the corresponding 
monodromy is
\be
M_\infty^{-k} M_{u_0} M_\infty^k = 
\pmatrix{1-4k&8k^2\cr -2& 1+4k\cr}=M(2k,1)
\ee
and similarly
\be
M_\infty^{-k} M_{-u_0} M_\infty^k
=\pmatrix{-1-4k&2+8k+8k^2\cr -2& 3+4k\cr} =M(2k+1,1)\ . 
\ee
So one sees that these monodromies correspond to dyons with $n_m=1$ and any
$n_e\in {\bf Z}$ becoming massless. Similarly one has e.g.
$M_{u_0}^{k} M_{-u_0} M_{u_0}^{-k}$ $=M(2k-1,-1)$, etc.

Let's come back to the question of how many singularities there are. 
Suppose there are
$p$ strong coupling singularities at $u_1, u_2, \ldots u_p$ in 
addition to the one-loop
perturbative singularity at  $u=\infty$. Then one has a factorisation 
analogous to
(\ref{cxii}):
\be\label{cxvii}
M_\infty =  M_{u_1}  M_{u_2} \ldots  M_{u_p}
\ee
with $M_{u_i}=M(n_m^{(i)}, n_e^{(i)})$ of the form (\ref{cxv}). 
It thus becomes a problem of
number theory to find out whether, for given $p$, there exist solutions to  
(\ref{cxvii}) with integer 
$n_m^{(i)}$ and $n_e^{(i)}$. For several low values of $p>2$ it has been 
checked
that there are no such solutions, and it seems likely that the same is 
true for all $p>2$.

\subsection{The solution}

Recall that our goal is to determine the exact non-perturbative 
low-energy effective action, i.e. determine the function $\cf(z)$ 
locally. This will be achieved, at least in principle, once we know 
the functions $a(u)$ and $a_D(u)$, since one then can invert the first 
to obtain $u(a)$, at least within a certain domain of the moduli 
space. Substituting this into $a_D(u)$ yields $a_D(a)$ which upon 
integration gives the desired $\cf(a)$.

So far we have seen that $\ad(u)$ and $a(u)$ are single-valued except for the
monodromies around $\infty, u_0$ and $-u_0$. As is well-known from complex 
analysis, this means that  $\ad(u)$ and $a(u)$ are really multi-valued 
functions with branch cuts, the branch points being  $\infty, u_0$ 
and $-u_0$. A typical example is 
$f(u)=\sqrt{u} F(a,b,c;u)$, where $F$ is the hypergeometric function. 
The latter has a
branch cut from $1$ to $\infty$. Similarly, $\sqrt{u}$ has a branch 
cut from $0$ to
$\infty$ (usually taken along the negative real axis), so that 
$f(u)$ has two branch
cuts joining the three singular points $0,1$ and $\infty$. When 
$u$ goes around any of
these singular points there is a non-trivial monodromy between 
$f(u)$ and one other
function $g(u)= u^d F(a',b',c';u)$. The three monodromy matrices 
are in (almost) one-to-one
correspondence with the pair of functions $f(u)$ and $g(u)$.

In the physical problem at hand one knows the monodromies, namely
\be\label{si}
M_{\infty}=\pmatrix{-1&2\cr 0&-1\cr}\ , \quad 
M_{u_0}=\pmatrix{1&0\cr -2&1\cr}\ , \quad
M_{-u_0}=\pmatrix{-1&2\cr -2&3\cr}
\ee
and one wants to determine the corresponding functions $\ad(u)$ and 
$a(u)$. As will be
explained, the monodromies fix $\ad(u)$ and $a(u)$ up to normalisation, 
which will be
determined from the known asymptotics (\ref{ci}) at infinity.

The precise location of $u_0$ depends on the renormalisation
conditions which can be chosen such that $u_0=1$. Assuming this choice 
in the
sequel will simplify somewhat the equations. If one wants to keep $u_0$, 
essentially
all one has to do is to replace $u\pm 1$ by 
${u\pm u_0\over u_0}={u\over u_0}\pm 1$.

\vskip 3.mm
\noindent
\underline{The differential equation approach}

\noindent
This approach to determining $\ad$ and $a$ was first exposed in \cite{AB}.
Monodromies typically arise from differential equations 
with periodic coefficients.
This is well-known in solid-state physics where one 
considers a Schr\"odinger equation
with a periodic potential\footnote{
The constant energy has been included into the potential, 
and the mass has been
normalised to $\half$.}
\be\label{sia}
\left[ -{\rmd^2\over \rmd x^2} + V(x)\right] \p(x)=0 
\quad , \quad V(x+2\pi)=V(x) \ .
\ee
There are two independent solutions $\p_1(x)$ and $\p_2(x)$. 
One wants to compare
solutions at $x$ and at $x+2\pi$. Since, due to the periodicity 
of the potential $V$,
the differential equation at $x+2\pi$ is exactly the same as at 
$x$, the set of
solutions must be the same. In other words, $\p_1(x+2\pi)$ and 
$\p_2(x+2\pi)$ must be
linear combinations of $\p_1(x)$ and $\p_2(x)$:
\be\label{sii}
\pmatrix{\p_1\cr \p_2\cr} (x+2\pi) = M \pmatrix{\p_1\cr \p_2\cr} (x) 
\ee
where $M$ is a (constant) monodromy matrix.

The same situation arises for differential equations in the complex plane with
meromorphic coefficients. Consider again the Schr\"odinger-type equation
\be\label{siii}
\left[ -{\rmd^2\over \rmd \xi^2} + V(\xi)\right] \p(\xi)=0
\ee
with meromorphic $V(\xi)$, having poles at $\xi_1, \ldots \xi_p$ and (in general) 
also at
$\infty$. The periodicity of the previous example is now replaced by the
single-valuedness of $V(\xi)$ as $\xi$ goes around any of the poles of $V$ 
(with $\xi-\xi_i$
corresponding roughly to $e^{ix}$). So, as $\xi$  goes once around any 
one of the $\xi_i$, the
differential equation (\ref{siii})  does not change. So by the same 
argument as above, the two solutions
$\p_1(\xi)$ and $\p_2(\xi)$, when continued along the path surrounding 
$\xi_i$ must again be
linear combinations of $\p_1(\xi)$ and $\p_2(\xi)$:
\be\label{siv}
\pmatrix{\p_1\cr \p_2\cr} \left(\xi+e^{2\pi i}(\xi-\xi_i)\right) 
= M_i \pmatrix{\p_1\cr \p_2\cr} (\xi) 
\ee
with a constant $2\times 2$-monodromy matrix $M_i$ for each of the 
poles of $V$. Of
course, one again has the factorisation condition (\ref{cxvii}) for 
$M_\infty$. It is
well-known, that non-trivial constant monodromies correspond to poles 
of $V$ that are
at most of second order. In the language of differential equations, 
(\ref{siii}) then only
has {\it regular} singular points.

In our physical problem, the {\it two} multivalued functions $\ad(\xi)$ 
and $a(\xi)$ have 3
singularities with non-trivial monodromies at $-1, +1$ and $\infty$. 
Hence they must be
solutions of a second-order differential equation (\ref{siii}) with 
the potential $V$ having
(at most) second-order poles precisely at these points. 
The general form of this potential is\footnote{
Additional terms in $V$ that naively look like first-order poles 
($\sim {1\over \xi-1}$ or ${1\over \xi+1}$)
cannot appear since they correspond to third-order poles at $\xi=\infty$.
}
\be\label{sv}
V(\xi)=-{1\over 4} \left[ {1-\l_1^2\over (\xi+1)^2} +  {1-\l_2^2\over (\xi-1)^2 }
-{1-\l_1^2-\l_2^2+\l_3^2\over (\xi+1)(\xi-1)} \right]
\ee
with double poles at $-1, +1$ and $\infty$. The corresponding residues are 
$-{1\over 4}(1-\l_1^2)$, $-{1\over 4}(1-\l_2^2)$ and $-{1\over 4}(1-\l_3^2)$. 
Without
loss of generality, I assume $\l_i\ge 0$. The corresponding differential 
equation
(\ref{siii}) is well-known in the mathematical literature 
since it can be transformed into the hypergeometric differential equation. The
transformation to the standard hypergeometric equation is readily performed 
by setting
\be\label{svi}
\p(\xi)=(\xi+1)^{\half (1-\l_1)} (\xi-1)^{\half (1-\l_2)}\,  
f\left( {\xi+1\over 2}\right) \ .
\ee
One then finds that $f$ satisfies the hypergeometric 
differential equation
\be\label{sviii}
x(1-x) f''(x)+[c-(a+b+1)x]f'(x)-abf(x)=0
\ee
with
\be\label{six}
a=\half (1-\l_1-\l_2+\l_3)\ , \quad
b=\half (1-\l_1-\l_2-\l_3)\ , \quad
c=1-\l_1 \ . 
\ee
The solutions of the hypergeometric equation (\ref{sviii}) can be 
written in many different
 ways due to the various identities between the hypergeometric 
 function $F(a,b,c;x)$
and products with powers, e.g. $(1-x)^{c-a-b} F(c-a,c-b,c;x)$, 
etc. A convenient choice for the two
independent solutions is the following 
\be\label{sx}
\begin{array}{rcl}
f_1(x)&=&(-x)^{-a}F(a,a+1-c,a+1-b;{1\over x})\crbig
f_2(x)&=&(1-x)^{c-a-b} F(c-a,c-b,c+1-a-b;1-x) \ . \crbig
\end{array}
\ee
$f_1$ and $f_2$ correspond to Kummer's solutions denoted $u_3$ 
and $u_6$ \cite{ERD}. The choice of
$f_1$ and $f_2$ is motivated by the fact that $f_1$ has simple 
monodromy properties
 around
$x=\infty$ (i.e. $\xi=\infty$) and $f_2$ has simple monodromy properties
around $x=1$ (i.e. $\xi=1$),
so they are good candidates to be identified with $a(\xi)$ and $\ad(\xi)$.

One can extract a great deal of information from the asymptotic 
forms of $\ad(\xi)$ and $a(\xi)$. As $\xi\to\infty$ one has 
$V(\xi)\sim -{1\over 4} \, {1-\l_3^2\over \xi^2}$, so that
the two independent solutions behave asymptotically as 
$\xi^{\half (1\pm \l_3)}$ if $\l_3
\ne 0$, and as $\sqrt{\xi}$ and $\sqrt{\xi}\ln \xi$ if $\l_3=0$. 
Comparing with (\ref{civ}) 
(with $u\to \xi$) we see
that the latter case is realised. Similarly, with $\l_3=0$, 
as $\xi\to 1$, one has $V(\xi)\sim -{1\over 4} 
\left(  {1-\l_2^2\over (\xi-1)^2}-{1-\l_1^2-\l_2^2\over 2(\xi-1)}
\right)$, where I have kept the subleading term. From the 
logarithmic asymptotics
(\ref{cx})
one then concludes $\l_2=1$ (and from the subleading term also 
$-{\l_1^2\over 8}={i\over \pi}{c_0\over a_0}$). The 
${\bf Z}_2$-symmetry ($\xi\to -\xi$) on the moduli
space then implies that, as $\xi\to -1$, the potential  
$V$ does not  have a double pole
either, so that also $\l_1=1$. Hence we conclude
\be\label{sxi}
\l_1=\l_2=1\ ,\ \ \l_3=0 \ \Rightarrow \ V(\xi)
= -{1\over 4}\, {1\over (\xi+1)(\xi-1)}
\ee
and $a=b=-\half,\ c=0$. Thus from (\ref{svi}) 
one has $\p_{1,2}(\xi)=f_{1,2}\left({\xi+1\over
2}\right)$. One can then verify that the two solutions
\be\label{sxiii}
\begin{array}{rcl}
\ad(u)&=&i \p_2(u)=i{u-1\over 2} F\left({1\over 2},{1\over 2},2;{1-u\over
2}\right)\crbig
a(u)&=&-2i \p_1(u)=\sqrt{2} (u+1)^{1\over 2} 
F\left(-{1\over 2},{1\over 2},1;{2\over u+1}\right)\crbig
\end{array}
\ee
indeed have the required monodromies (\ref{si}), as well 
as the correct asymptotics.

It might look as if we have not used the monodromy properties 
to determine $\ad$ and
$a$ and that they have been determined only from the asymptotics. 
This is not entirely
true, of course. The very fact that there are non-trivial monodromies 
only at $\infty,
+1$ and $-1$ implied that $\ad$ and $a$ must satisfy the second-order 
differential
equation (\ref{siii}) with the potential (\ref{sv}). 
To determine the $\l_i$ we then used the
asymptotics of $\ad$ and $a$. But this is (almost) the same as using 
the monodromies
since the
latter were obtained from the asymptotics.

Using the integral representation of the hypergeometric function, the solution
(\ref{sxiii}) can be nicely rewritten as 
\be\label{sxiv}
\ad(u)={\sqrt{2}\over \pi} 
\int_1^u {\rmd x\ \sqrt{x-u}\over \sqrt{x^2-1} }\ , \quad
a(u)={\sqrt{2}\over \pi} \int_{-1}^1 {\rmd x\ \sqrt{x-u}\over \sqrt{x^2-1} }\ . 
\ee

One can invert the second equation (\ref{sxiii})  to obtain $u(a)$, within a 
certain domain, and 
insert the result into $\ad(u)$ to obtain $\ad(a)$. Integrating
with respect to $a$ yields $\cf(a)$ and hence the low-energy effective action. 
I should
stress that this expression for $\cf(a)$ is not globally
valid but only on a certain portion of the moduli space. Different analytic
continuations must be used on other portions.

\vskip 3.mm
\noindent
\underline{The approach using elliptic curves}

\noindent
In their paper \cite{SW}, Seiberg and Witten do not use the 
differential equation approach just
described, but rather introduce an auxiliary construction: a certain
 elliptic curve by
means of which two functions with the correct monodromy properties are 
constructed. I
will not go into details here, but simply sketch this approach.

To motivate their construction {\it a posteriori}, we notice the 
following: from the
integral representation (\ref{sxiv}) it is natural to consider the 
complex $x$-plane. More
precisely, the integrand has square-root branch cuts with branch 
points at $+1, -1, u$
and $\infty$. The two branch cuts can be taken to run from $-1$ to 
$+1$ and from $u$ to
$\infty$. The Riemann surface of the integrand is two-sheeted with 
the two sheets
connected through the cuts. If one adds the point at infinity to each 
of the two
sheets, the topology of the Riemann surface is that of two spheres 
connected by two
tubes (the cuts), i.e. a torus. So one sees that the Riemann
surface of the integrand in (\ref{sxiv}) has genus one. This is 
the elliptic curve considered
by Seiberg and Witten.

As is well-known, on a torus there are two independent non-trivial closed paths
(cycles). One cycle ($\g_2$) can be taken to go once around the cut 
$(-1,1)$, and the
other cycle ($\g_1$) to go from $1$ to $u$ on the first sheet and 
back from $u$ to $1$
on the second sheet. The solutions $\ad(u)$ and $a(u)$ in (\ref{sxiv}) 
are precisely the
integrals of some suitable differential $\l$ along the two cycles $\g_1$ 
and $\g_2$:
\be\label{sxiva}
\ad=\oint_{\g_1} \l  \quad , \quad a=\oint_{\g_2} \l  \quad , \quad
\l={\sqrt{2}\over 2\pi} {\sqrt{x-u}\over \sqrt{x^2-1} } \rmd x \ .
\ee
These integrals are called period integrals. They are known to satisfy a
second-order differential equation, the so-called Picard-Fuchs equation, that is
nothing else than our Schr\"odinger-type equation (\ref{siii}) with $V$ 
given by (\ref{sxi}).

How do the monodromies appear in this formalism? As $u$ goes once 
around $+1,-1$ or
$\infty$, the cycles $\g_1, \g_2$ are changed into linear combinations 
of themselves
with integer coefficients:
\be\label{sxv}
\pmatrix{\g_1\cr \g_2} \to M \pmatrix{\g_1\cr \g_2} \quad , \quad
M\in Sl(2, {\bf Z}) \ .
\ee
This immediately implies
\be\label{sxvi}
\pmatrix{\ad\cr a} \to M \pmatrix{\ad\cr a}
\ee
with the same $M$ as in (\ref{sxv}). The advantage here is that one 
automatically gets
monodromies with {\it integer} coefficients. The other advantage is that
\be\label{sxvii}
\tau(u)={\rmd \ad/\rmd u\over  \rmd a/\rmd u}
\ee
can be easily seen to be the $\tau$-parameter describing the complex 
structure of the
torus, and as such is garanteed to satisfy
$\im\tau(u) >0$
which was the requirement for positivity of the metric on moduli space.

To motivate the appearance of the genus-one elliptic curve 
(i.e. the torus) {\it a
priori} - without knowing the solution (\ref{sxiv}) from the 
differential equation approach -
Seiberg and Witten remark that the three monodromies are all 
very special: they do not
generate all of $Sl(2, {\bf Z})$ but only a certain subgroup $\G(2)$ of 
matrices 
in
$Sl(2, {\bf Z})$ congruent to $1$ modulo $2$. Furthermore, they remark that the
$u$-plane with punctures at $1,-1,\infty$ can be thought of as the  
quotient of the
upper half plane $H$ by $\G(2)$, and that $H/\G(2)$ naturally 
parametrizes (i.e. is the
moduli space of) elliptic curves described by
\be\label{sxix}
y^2=(x^2-1)(x-u) \ .
\ee
Equation (\ref{sxix}) corresponds to the genus-one Riemann surface 
discussed above, and it is
then natural to introduce the cycles $\g_1, \g_2$ and the differential 
$\l$ from
(\ref{sxiv}).
The rest of the argument then goes as I just exposed.

\vskip 3.mm
\noindent
\underline{Summary}

\noindent
Let's summarise what we have learnt so far.
 We have seen realised a version of
electric-magnetic duality accompanied by a duality transformation on 
the expectation
value of the scalar (Higgs) field, $a\leftrightarrow \ad$. There is a 
manifold of
inequivalent vacua, the moduli space $\M$, corresponding to different 
Higgs expectation
values. The duality relates strong coupling regions in $\M$ to the 
perturbative region
of large $a$ where the effective low-energy action is known asymptotically
in terms of $\cf$. Thus duality allows us to determine the latter 
also at strong
coupling. The holomorphicity condition from ${\cal N}=2$ supersymmetry 
then puts such strong
constraints on $\cf(a)$, or equivalently on $\ad(u)$ and $a(u)$ 
that the full functions
can be determined solely from their asymptotic behaviour at the strong 
and weak coupling singularities of $\M$.

\newcommand{\Z}{{\bf Z}}
\newcommand{\R}{{\bf R}}
\newcommand{\C}{{$\cal C$}}
\newcommand{\sw}{{\cal S}_W}
\newcommand{\ssp}{{\cal S}_{S+}}
\newcommand{\ssm}{{\cal S}_{S-}}
\newcommand{\sso}{{\cal S}_{S0}}
\newcommand{\rw}{{\cal R}_W}
\newcommand{\rsm}{{\cal R}_{S-}}
\newcommand{\rsp}{{\cal R}_{S+}}
\newcommand{\rso}{{\cal R}_{S0}}

\section{The spectrum of stable BPS states: pure ${\rm SU}(2)$ 
without hypermultiplets}

Knowing the low-energy effective action of the ${\cal N}=2$ gauge 
theory allows us to study the dynamics of the light degrees of freedom. 
This certainly is quite an achievement. One may want to go further, 
however. The heavy, massive fields all must be BPS states since 
otherwise the multiplets contain spins exceeding one. Hence they must 
satisfy the BPS bound relating their masses to their charges. Studying 
their detailed dynamics is a difficult problem, in most cases well 
beyond what can be done. It is already a non-trivial question to 
study their existence and stability as the effective 
coupling changes, i.e. as one 
moves around in the moduli space. This problem though has been solved 
with somewhat surprising results. In this section I will review this 
solution in the simplest case corresponding to the theory studied in 
the previous section: gauge group ${\rm SU}(2)$ and no elementary
hypermultiplets. 
In the next section, I will review the results for the more involved 
cases where massless or massive elementary hypermultiplets are present.

\subsection{BPS states, charge lattice and curve of marginal stability}

Recall that the ${\cal N}=2$ susy algebra has long and short representations 
and for short representations (BPS states) the BPS bound must be satisfied:
\be\label{bpsbound}
m=\sqrt{2}\vert Z \vert = \sqrt{2}\vert   n_e a(u) - n_m a_D(u)\vert \ .
\ee
We have seen that the central charge 
$Z$ can be written in terms of the standard symplectic invariant
$\eta(p,\Omega)$ of $p=(n_e, n_m)$ and $\Omega=(a_D,a)$ which is such that 
$\eta(G p,G \Omega)=\eta(p,\Omega)$ for any $G\in Sp(2,\Z)\equiv SL(2,\Z)$.

For {\it fixed} quantum numbers $(n_e,n_m)$ a BPS state has the minimal mass 
and must be stable. The question then is whether it can decay into two 
(or more) other states such that the charge quantum numbers are conserved. 
Take the example of a dyon with $n_e=1$ and $n_m=1$. By charge conservation 
this could decay into a monopole $(n_e,n_m)=(0,1)$ and a W-boson 
$(n_e,n_m)=(1,0)$. Kinematically however this is impossible, since the 
sum of the masses of the latter is larger than the mass of the initial 
dyon. To discuss the general case one draws the charge lattice in the 
complex plane as follows. Generically, for a given
point $u$ in moduli space, $a$ and $a_D$ are two complex numbers 
such that $a_D/a\notin \R$ and all possible central charges form a 
lattice in the complex plane generated by $a$ and $(-\ad)$, see Fig. 1.
\begin{figure}[htb]
\vspace{9pt}
\centerline{ \fig{4.5cm}{4.5cm}{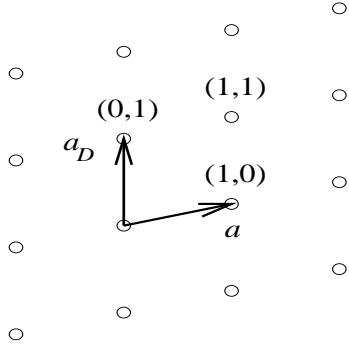} }
\caption{The lattice of central charges for generic $a_D$ and $a$}
\label{lattice}
\end{figure}
Each lattice point corresponds to an a priori possible BPS state 
$(n_e, n_m)$ whose mass is simply
its euclidean distance from the origin. For the above example of the 
$(1,1)$ dyon it is clear by the triangle inequality of elementary
geometry, that the sum of the masses of the decay products $(1,0)$ and $(0,1)$
would be larger than the mass of the $(1,1)$ dyon. 
Obviously, decay into non-BPS states is even more impossible since they
would have even larger masses for the same charge quantum numbers.
Hence the $(1,1)$ dyon is stable.
The same argument applies to all BPS states $(n_e, n_m)$ such that 
$(n_e, n_m)\ne q(n,m)$ with $n,m,q\in\Z$, $q\ne \pm 1$:
states with $n_e$ and $n_m$ relatively prime are stable.

The preceeding argument fails if 
\be\label{ratio}
w(u)\equiv {a_D(u)\over a(u)} \in \R\ ,
\ee 
since then the 
lattice collapses onto a single line and decays of otherwise stable 
BPS states become possible. It is thus of interest to determine  
the set of all such $u$, i.e.
\be\label{margcurve}
{\cal C}=\{u \in {\bf C}\  \vert\  w(u)\equiv {a_D(u)\over a(u)} \in \R\}\ ,
\ee
which is called the curve 
of marginal stability \cite{SW,ARG}. Given the explicit form of 
$a_D(u)$ and $a(u)$ it is straightforward to determine \C\ numerically 
\cite{FB}, see Fig. 2, although it can also be done analytically \cite{MATO}. 
\begin{figure}[htb]
\vspace{9pt}
\centerline{ \fig{6cm}{4cm}{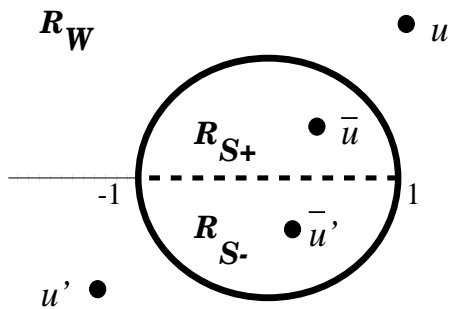} }
\caption{In the $u$ plane, we show the curve \C\ of marginal stability 
which is almost an ellipse centered at the origin (thick line), 
the cuts of $a(u)$ and $a_D(u)$ (dotted and dashed lines), as well 
as the definitions of the weak-coupling region $\rw$ and the
strong-coupling region ($\rsp \cup \rsm$).}
\label{curve}
\end{figure}
The precise form of the curve however is irrelevant for our
purposes. What is important is that as $u$ varies along the curve, 
$a_D/a$ takes all values in $[-1,1]$. More precisely, if we call 
${\cal C}^\pm$ the parts of \C\ in the
upper and lower half $u$ plane, then
\begin{eqnarray}
{a_D\over a}(u) &\in& [-1,0] \quad {\rm for}\ u\in {\cal C}^+ \ , \cr
&{}& \cr
{a_D\over a}(u) &\in& [0,1] \quad {\rm for}\ u\in {\cal C}^- \ ,
\label{iv}
\end{eqnarray}
with the value being discontinuous at $u=-1$ due to the cuts of 
$a_D$ and $a$ running along the real axis from $-\infty$ to $+1$.
More precisely, ${\ad\over a}(u)$ increases monotonically from 
$-1$ at $u=-1+i\e$ to $+1$ at $u=-1-i\e$ as one follows the curve 
clockwise. Obviously ${\ad\over a}=0$ at $u=1$.

The curve \C\ separates the moduli space into two distinct regions: inside the curve
and outside the curve, see Fig. 2. If two points $u$ and $u'$ are in the same region,
i.e. if they can be joined by a path not crossing \C\ then the spectrum of BPS states
(by which we mean the set of quantum numbers $(n_e, n_m)$ that do exist) is
necessarily the same at $u$ and $u'$. Indeed, start with a given stable BPS state at
$u$. Then imagine deforming the theory adiabatically so that the scalar field $\f$
slowly changes its vacuum expectation value and $\langle {\rm tr \, }\f^2\rangle$
moves from $u$ to $u'$. In doing so, the BPS state will remain stable and it cannot
decay at any point on the path. Hence it will also exist at $u'$. If, however, $u$ and
$\tilde u$ are in different regions so that the path joining them must cross the curve
\C\ somewhere, then the initial BPS state will no longer be stable as one crosses the
curve and it can decay. Hence the spectrum at $u$ and $\tilde u$ need not be the same.

As an example, consider the possible decay of the W boson $(1,0)$ when crossing the
curve on ${\cal C}^+$ at a point where $a_D/a =r$ with $r$ any real number between
$-1$ and $0$. Charge conservation alone allows for the reaction
\begin{equation}
(1,0)\to (1,-1) + (0,1) \ .
\label{v}
\end{equation}
On ${\cal C}^+$, and only on ${\cal C}^+$, we also have the equality of masses, thanks
to
\begin{eqnarray}
\vert a+a_D\vert + \vert a_D\vert &=&\vert a \vert \left( \vert 1+r\vert +\vert r \vert
\right)  \cr
&=& \vert a \vert \left( 1+r -r\right) = \vert a \vert  \ .
\label{vi}
\end{eqnarray}
Had one crossed the curve in the lower half plane instead, $r$ would have been between
$0$ and $+1$ and the dyon $(1,-1)$ would have been 
decribed as $(1,1)$ (see below), and eq.
(\ref{vi}) would have worked out correspondingly.

Since the region of moduli space outside the curve contains the semi-classical domain
$u\to\infty$, we refer to this region as the semi-classical or weak-coupling region
${\cal R}_W$ and to the region inside the curve as the strong-coupling region
${\cal R}_S$. We call the
corresponding spectra also weak and strong-coupling spectra $\sw$ and ${\cal S}_S$.
This terminology is used due to the above-explained continuity of the spectra
throughout each of the two regions. Nevertheless, the physics close to the curve is
always strongly coupled even in the so-called weak-coupling region.

\subsection{The main argument and the weak-coupling spectrum}

The important property of the curve \C\ of marginal stability is

\noindent 
$\bullet$
{\bf P1} : Massless states can only occur on the curve \C.

\noindent
The proof is trivial: If we have a massless state at some point $u$, it necessarily is
a BPS state, hence $m(u) = 0$ implies $n_e a(u) - n_m a_D(u) = 0$ which can be
rewritten as $(a_D/a)(u) = n_e/n_m$. But $n_e/n_m$ is a real number, hence 
$(a_D/a)(u)$ is real, and thus $u\in$ \C. Indeed the points $u=\pm 1$ where the
magnetic monopole and the dyon $(\pm 1,1)$ become massless are on the curve. The
converse statement obviously also is true:

\noindent 
$\bullet$
{\bf P2} : A BPS state $(n_e, n_m)$ with $n_e/n_m \in [-1,1]$ becomes
massless somewhere on the curve \C.

\noindent
Of course, it will become massless precisely at the point $u\in$ \C\ where 
$(a_D/a)(u) = n_e/n_m$. Strictly speaking, in its simple form, this only 
applies to
BPS states in the weak-coupling region, since the description of BPS 
states in the
strong-coupling region is slightly more involved as shown below. Let 
me now state the
main hypothesis.

\noindent 
$\bullet$
{\bf H} : A state becoming massless always leads to a singularity of the
low-energy effective action, and hence of $\left( a_D(u),\ a(u)\right)$. The
Seiberg-Witten solution for $\left( a_D(u),\ a(u)\right)$ is correct and 
there are only two singularities at finite $u$, namely $u=\pm 1$.

\noindent
Then the argument we will repeatedly use goes like this: If a certain 
state would
become massless at some point $u$ on moduli space, it would lead to an 
extra
singularity which we know cannot exist. Hence this state either is the 
magnetic
monopole $\pm (0,1)$ or the $\pm (\pm 1,1)$ dyon and $u=\pm 1$, or 
this state cannot
exist.

As an immediate consequence we can show that the weak-coupling 
spectrum cannot
contain BPS states with $\vert n_m\vert > \vert n_e\vert >0$. 
Indeed, for such a state,
$-1 < n_e/n_m < 1$ and it would be massless at the point $u$ 
on \C\ where 
$(a_D/a)(u) = n_e/n_m$. Since  $\vert n_m\vert > \vert n_e\vert >0$ 
it is neither the
monopole ($n_e=0$) nor the $(\pm 1,1)$ dyon, hence it cannot exist.

To determine which states are in $\sw$ one uses a global symmetry. 
Taking $u\to
e^{2\pi i} u$ along a path outside \C\ does not change the theory 
since one comes back
to the same point of moduli space, and hence must leave $\sw$ 
invariant. But it
induces a monodromy transformation
\begin{equation}
\pmatrix{ n_e\cr n_m\cr} \, \to \, M_\infty \pmatrix{ n_e\cr n_m\cr} \, , \,
M_\infty=\pmatrix{ -1&\hfill 2\cr \hfill 0 & -1\cr} .
\label{vii}
\end{equation}
In other words, $M_\infty \sw = \sw$. Now, we know that $\sw$ 
contains at least the
two states that are responsible for the singularities, namely 
$(0,1)$ and $(1,1)$
together with their antiparticles $(0,-1)$ and $(-1,-1)$. 
Applying $M_\infty^{\pm 1}$ on these
two states generates all dyons $(n,\pm 1),\ n\in \Z$. 
This was already clear from \cite{SW}. But
now we can just as easily show that there are no other dyons in 
the weak-coupling
spectrum. If there were such a state $\pm (k,m)$ with 
$\vert m\vert \ge 2$, then applying
$M_\infty^n,\ n\in \Z$, there would also be all states 
$\pm (k-2n m,m)$. The latter
would become massless somewhere on \C\ if 
$(k-2n m/m)=(k/m)-2n \in [-1,1]$. Since there
is always such an $n\in \Z$, this state, and 
hence $\pm (k,m)$ cannot exist in $\sw$.
Finally. the W boson which is part of the 
perturbative spectrum is left invariant by
$M_\infty$: $M_\infty (1,0) = - (1,0)$, where 
the minus sign simply corresponds to the
antiparticle. Hence we conclude
\begin{equation}
\sw = \left\{ \pm (1,0),\  \pm (n,1),\ n\in \Z \right\} \ .
\label{viii}
\end{equation}
This result was already known from semi-classical considerations 
on the moduli space
of multi-monopole configurations \cite{SEN,STERN}, but it is nice 
to rederive it in
this particularly simple way. Now let us turn to the new results of
\cite{FB} concerning the
strong-coupling spectrum.

\subsection{The $\Z_2$ symmetry}

As discussed above, the classical susy $SU(2)$ Yang-Mills theory 
has a $U(1)_R$ $R$-symmetry acting on the
scalar $\f$ as $\f\to e^{2 i \alpha}\f$ so that $\f$ has charge two. 
In the quantum
theory this global symmetry is anomalous, and it is easy to see from 
the explicit form
of the one-loop and instanton contributions to the low-energy effective
action (i.e.
to $\F$) that only a discrete subgroup $\Z_8$ survives, corresponding 
to phases
$\alpha={2\pi \over 8} k,\ k\in\Z$. Hence under this $\Z_8$ one has 
$\f^2\to (-)^k
\f^2$. This $\Z_8$ is a symmetry of the quantum action and of the 
Hamiltonian, but a
given vacuum with $u=\langle {\rm tr \, }\f^2\rangle \ne 0$ is 
invariant only under
the $\Z_4$ subgroup corresponding to even $k$. The quotient 
(odd $k$) is a $\Z_2$
acting as $u\to -u$. Although a given vacuum breaks the full 
$\Z_8$ symmetry, the
broken symmetry (the $\Z_2$) relates physically equivalent 
but distinct vacua. In
particular, the mass spectra at $u$ and at $-u$ must be the 
same. This means that for
every BPS state $(n_e, n_m)$ that exists at $u$ there must 
be some BPS state $(\tilde
n_e, \tilde n_m)$ at $-u$ having the same mass:
\begin{equation}
\vert \tilde n_e a(-u) - \tilde n_m a_D(-u) \vert 
= \vert n_e a(u) - n_m a_D(u) \vert \ .
\label{ix}
\end{equation}
This equality shows that there must exist a matrix
$G\in Sp(2,\Z)$ 
such that
\begin{eqnarray}
\pmatrix{ \tilde n_e \cr \tilde n_m \cr}
&=&\pm G \pmatrix{ n_e\cr n_m\cr} \ , \cr
&{}& \cr
\pmatrix{ a_D\cr a\cr} (-u) &=& e^{i\omega}\,  G \pmatrix{ a_D\cr a\cr} (u)
\label{x}
\end{eqnarray}
where $e^{i\omega}$ is some phase. Indeed, from the explicit expressions 
of $a_D$ and 
$a$ one finds, using standard relations between hypergeometric functions, that
\begin{equation}
G= G_{W,\epsilon} \equiv \pmatrix{ 1&\epsilon\cr 0& 1\cr}\, ,\ e^{i\omega}=
e^{-i\pi\epsilon/2}
\label{xi}
\end{equation}
where $\epsilon =\pm 1$ according to whether $u$ is in the upper or lower half plane.
The subscript $W$ indicates that this is the matrix to be used in the weak-coupling
region, while for the strong-coupling region there is a slight subtlety to be
discussed soon. We have just shown that  for any BPS state $(n_e, n_m)$ existing at
$u$ (in the weak-coupling region) with mass $m$ there exists another BPS state
$(\tilde n_e, \tilde n_m)=\pm G_{W,\epsilon} (n_e, n_m)$ at $-u$ with the same mass
$m$. Now, since both $u$ and $-u$ are outside the curve \C, they can be joined by a
path never crossing \C, see Fig. 3, and hence the BPS state $(\tilde n_e, \tilde n_m)$ must also
exist at $u$, although with a different mass $\tilde m$. So we have been able to use
the broken symmetry to infer the existence of the state $(\tilde n_e, \tilde n_m)$ at
$u$ from the existence of  $(n_e, n_m)$ at the {\it same} point $u$ of moduli space.
Starting from the magnetic monopole $(0,1)$ at $u$ in the upper half plane (outside
\C) one deduces the existence of all dyons $(n,1)$ with $n\ge 0$. Taking similarly $u$
in the lower half plane (again outside \C) one gets all dyons $(n,1)$ with $n\le 0$.
The W boson $(1,0)$ is invariant under $G_{W,\epsilon}$. Once again, one generates
exactly the weak-coupling spectrum $\sw$ of (\ref{viii}), and clearly $G_{W,\epsilon}
\sw = \sw$.

\begin{figure}[htb]
\vspace{9pt}
\centerline{ \fig{5.5cm}{4.5cm}{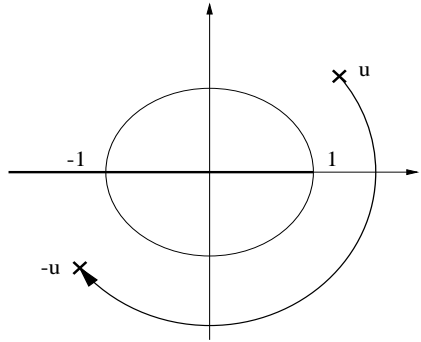} }
\caption{Taking $u$ to $u'=-u$ in the weak-coupling region ${\cal R}_W$
without crossing the
cuts on $(-\infty,1]$}
\label{weakcoupldeform}
\end{figure}

\subsection{The strong-coupling spectrum}

It is in the strong-coupling region that this $\Z_2$ symmetry will show its full
power. Here $M_\infty$ no longer is a symmetry, since a monodromy circuit around
infinity can be deformed all through the weak-coupling region but it cannot cross \C\
into the strong-coupling region since the state that is taken along this circuit may
well decay upon crossing the curve \C. The relations (\ref{x}) and (\ref{xi}) 
expressing $a_D(-u),
a(-u)$ in terms of $a_D(u), a(u)$ nevertheless remain true. What needs to be
reexamined is the relation between $\tilde n_e, \tilde n_m$ and $n_e, n_m$. This is
due to the fact that there is a cut of the function $a(u)$ running between $-1$ and
$1$, separating the strong-coupling region ${\cal R}_S$ into two parts, 
$\rsp$ and $\rsm$, 
as shown in Fig. 2.
As a consequence, the same BPS state is described by two different sets of
integers in $\rsp$ and $\rsm$. If we call the corresponding spectra $\ssp$ and $\ssm$
then we have
\begin{eqnarray}
\ssm&=&M_1^{-1} \ssp\ ,\ \pmatrix{n_e'\cr n_m'\cr} = M_1^{-1} \pmatrix{n_e\cr n_m\cr}
\ ,\cr
M_1^{-1}&=&\pmatrix{1&0\cr 2&1\cr} \ .
\label{xia}
\end{eqnarray}
This change of description is easily explained: take a BPS state $(n_e, n_m)\in \ssp$
at a point $u\in \rsp$ and transport it to a point $u'\in \rsm$, see Fig. 4. 
In doing so, its
mass varies continuously and nothing dramatic can happen since one does not cross the
curve \C. Hence, as one crosses from $\rsp$ into $\rsm$, 
the functions $a_D$ and $a$ must also vary
smoothly, which means that at $u'\in\rsm$ one has the analytic continuation of
$a_D(u)$ and $a(u)$. But this is not what one calls $a_D$ and $a$ in $\rsm$. Rather,
these analytic continuations $\tilde a_D(u')$ and $\tilde a(u')$ are related to
$a_D(u')$ and $a(u')$ by the monodromy matrix around $u=1$ which is $M_1$ as
\begin{equation}
\pmatrix{\widetilde {a_D}(u')\cr \tilde a(u')\cr} = 
M_1 \pmatrix{ a_D(u')\cr a(u')\cr} \ .
\label{xib}
\end{equation}
Hence the
mass of the BPS state at $u'$ is $\sqrt{2} \vert n_e \tilde a(u') - n_m \tilde
a_D(u')\vert$ $= \sqrt{2}\vert n_e' a(u') - n_m' a_D(u') \vert$ where $n_e',\, n_m'$
are given by eq. (\ref{xia}). 

\begin{figure}[htb]
\vspace{9pt}
\centerline{ \fig{5.5cm}{4.5cm}{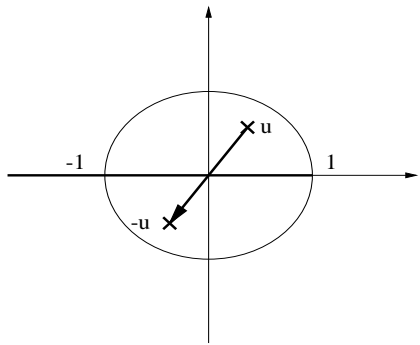} }
\caption{Taking $u$ to $u'=-u$ inside the strong-coupling 
region ${\cal R}_S$ one has to cross the cut on $[-1,1]$.}
\label{strongcoupldeform}
\end{figure}

As a consequence of the two different descriptions of
the same BPS state, the $G$-matrix implementing the $\Z_2$ transformation on the
spectrum has to be modified. As before, from 
the existence of $(n_e, n_m)$ at $u\in \rsp$ one
concludes the existence of a state $G_{W,+} (n_e, n_m)$ at $-u\in \rsm$. This
same state must then also exist at $u$ but is described as 
$M_1 G_{W,+} (n_e, n_m)$. Had one started with a $u\in \rsm$ the relevant matrix would
have been $M_1^{-1} G_{W,-}$. Hence, in the strong-coupling region $G_{W,\pm}$ is
replaced by
\begin{equation}
G_{S,\epsilon}=(M_1)^\epsilon G_{W,\epsilon} = \pmatrix{\hfill 1&\hfill\epsilon\cr
-2\epsilon&-1\cr}\ ,
\label{xii}
\end{equation}
and again one concludes that the existence of a BPS state $(n_e, n_m)$ at $u\in {\cal
R}_{S,\epsilon}$ implies the existence of another BPS state $G_{S,\epsilon} (n_e,
n_m)$ at the {\it same} point $u$. The important difference now is that
\be\label{gsquareisone}
G_{S,\epsilon}^2=-{\bf 1} \ ,
\ee 
so that applying this argument twice just gives back
$(-n_e, -n_m)$. But this is the antiparticle of $(n_e, n_m)$ and always exists
together with $(n_e, n_m)$. As far as the determination of the spectrum is concerned
we do not really need to distinguish particles and antiparticles. In this sense,
applying $G_{S,\epsilon}$ twice gives back the same BPS state. Hence in the
strong-coupling region, all BPS states come in pairs, or $\Z_2$ doublets (or quartets
if one counts particles and antiparticles separately):
\be\label{xiii}
\pm \pmatrix{n_e\cr n_m\cr}\in \ssp \, \Leftrightarrow \, \pm\,  G_{S,+} 
\pmatrix{n_e\cr n_m\cr} 
=\pm \pmatrix{ n_e+n_m\cr -2n_e-n_m\cr} \in \ssp 
\ee
and similarly for $\ssm$.
An example of such a doublet is the magnetic monopole $(0,1)$ and the dyon
$(1,-1)=-(-1,1)$ which are the two states becoming massless at the $\Z_2$-related
points $u=1$ and $u=-1$. Note that in $\ssm$ the monopole is still described as
$(0,1)$ while the same dyon is described as $(1,1)$. It is now easy to show that this
is the only doublet one can have in the strong-coupling spectrum. Indeed, one readily
sees that either $n_e/n_m \equiv r$ is in $[-1,0]$ or $(n_e+n_m)/(-2 n_e-n_m)= -
(r+1)/(2r+1)$ is in $[-1,0]$. This means that one or the other member of the $\Z_2$
doublet (\ref{xiii}) becomes massless somewhere on ${\cal C}^+$, the part of the curve
\C\ that can be reached from $\rsp$. But as already repeatedly argued, the only states
ever becoming massless are the magnetic monopole $(0,1)$ and the dyon $(1,-1)$. Hence
no other $\Z_2$ doublet can exist in the strong-coupling spectrum and we conclude that
\be\label{xiv}
\ssp = \left\{ \pm (0,1), \pm (-1,1) \right\} \quad
\Leftrightarrow \quad
\ssm = \left\{ \pm (0,1), \pm (1,1) \right\} \ .
\ee
$\bullet$
{\bf P3} : The strong-coupling spectrum consists of only those BPS states that are
responsible for the singularities. All other weak-coupling, i.e. semi-classical BPS
states must and do decay consistently into them when crossing the curve \C.

\noindent
We have shown above the example of the decay of the 
W boson, cf. eq. (\ref{v}), but it is just as
simple to show consistency of the other decays \cite{FB}.

When adding massless quark hypermultiplets next, we will see that the details of the
spectrum change, however, the conclusion P3 will remain the same.

\section{Generalisation to ${\cal N}=2$ susy QCD : including hypermultiplets}

\subsection{Massless hypermultiplets}

We will continue to consider only the gauge group SU(2) 
as studied in \cite{SWII}. First, in this subsection, we
will also restrict ourselves to the case of vanishing bare masses 
of the quark hypermultiplets. The number of hypermultiplets is usually
referred to as the number of flavours $N_f$. Although these quark 
hypermultiplets have no bare masses in the original Lagrangian, they 
get physical masses through the Higgs mechanism much like the W-bosons. 
These masses are given by the same BPS mass formula as in the previous 
section.

We will be very qualitative and describe only the results,
referring the reader to \cite{BF} for details. The main difference with respect to the
previous case of pure Yang-Mills theory is that now the BPS states carry
representations of the flavour group which is the covering group of $SO(2N_f)$, namely
$SO(2)$ for one flavour, $Spin(4)=SU(2)\times SU(2)$ for two flavours, and
$Spin(6)=SU(4)$ for three flavours. We will present each of the three cases separately.
In all cases:\\
$\bullet$ There is a curve of marginal stability diffeomorphic to a circle 
and going through all (finite) singular points of moduli space.  \\
$\bullet$ The BPS spectra are discontinuous across these curves.  \\
$\bullet$ The strong-coupling spectra (inside the curves) contain only 
those BPS states that can become massless and are responsible for the 
singularities. They form a multiplet (with different 
masses) of the broken global discrete symmetry,  except for $N_f=3$ 
where there is no such symmetry.   \\
$\bullet$ All other semi-classical BPS states must and do decay 
consistently when crossing the curves. \\
$\bullet$ The weak-coupling, i.e. semi-classical BPS spectra, contain 
no magnetic charges larger than one for $N_f=0,1,2$ and no magnetic charges 
larger than two for $N_f=3$.

It is useful to slightly change conventions for $a(u)$ and 
$n_e$: we henceforth replace
\be\label{newconvent}
a(u) \to a(u)/2 \quad , \quad n_e\to 2 n_e \ ,
\ee
so that the W-bosons now have $(n_e,n_m)=(2,0)$. 
This is useful since the hypermultiplets correspond to ``quarks" in the fundamental 
representation of ${\rm SU}(2)$ and hence have half the charge 
of the gauge bosons which are in the adjoint. With the new conventions the 
``quarks" have integer rather than half-integer charges. 
Also, the spectra of the pure gauge theory obtained above now read:
\begin{eqnarray}
\sw&=& \left\{ \pm (2,0),\  \pm (2n,1),\ n\in \Z \right\} \crbig
\ssp &=& \left\{ \pm (0,1), \pm (-2,1) \right\} 
\Leftrightarrow
\ssm = \left\{ \pm (0,1), \pm (2,1) \right\} \ .
\label{hxiv}
\end{eqnarray}

\noindent
\underline{$N_f=1$}

\noindent
According to Seiberg and Witten \cite{SWII} there are 3 singularities at finite points
of the Coulomb branch of the moduli space. They are related by a global discrete $\Z_3$
symmetry. This $\Z_3$ is the analogue of the $\Z_2$ symmetry discussed previously. Its
origin is slightly more complicated, however, since the original $\Z_{12}$ is due to a
combination of a $\Z_6$ coming from the anomalous $U(1)_R$ symmetry and of the
anomalous flavour-parity of the $O(2N_f)$ flavour group. In any case, the global
discrete symmetry of the quantum theory is $\Z_{12}$. The vacuum with non-vanishing
value of $u=\langle {\rm tr\, }\f^2 \rangle$ breaks this to $\Z_4$. The quotient $\Z_3$
acting as $u\to e^{\pm 2\pi i/3} u$ then is a symmetry relating different but
physically equivalent vacua. The three singular points are due to a massless
monopole $(0,1)$, a massless dyon $(-1,1)$ and another massless dyon $(-2,1)$. Again
there is a curve of marginal stability that was obtained from the explicit expressions
for $a_D(u)$ and $a(u)$ \cite{BF}. It is almost a circle, and of course, it goes
through the three singular points, see Fig. 5, 
\begin{figure}[htb]
\vspace{9pt}
\centerline{ \fig{5.5cm}{4.5cm}{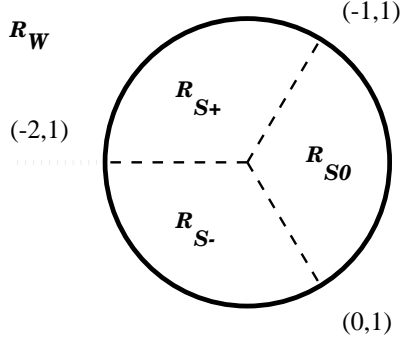} }
\caption{The curve of marginal stability and the three different portions of the
strong-coupling region separated by the cuts, for $N_f=1$}
\label{matter}
\end{figure}
where we also indicated the various cuts
and correspondingly different portions $\rsp, \rsm, \rso$ of the
strong-coupling region ${\cal R}_S$. So here one 
needs to introduce three different
desciptions of the same strong-coupling BPS state.
The corresponding spectra are denoted ${\cal S}_{S+}$,
${\cal S}_{S0}$ and ${\cal S}_{S-}$. 
We will not give $\ad(u)$ and $a(u)$ explicitly here, 
but refer the reader to ref \cite{BF}.
The ratio $a_D/a$ increases
monotonically from $-2$ to $+1$ as one goes along the curve in 
a clockwise sense,
starting at the point where $(-2,1)$ is massless. Then using 
exactly the same type of
arguments as we did before, one obtains the weak and 
strong-coupling spectra. All
states in the latter now belong to a single $\Z_3$ triplet, 
containing precisely the
three states responsible for the singularities. Denoting a 
BPS state by $(n_e, n_m)_S$
where $S$ is the $SO(2)$ flavour charge, and denoting its 
antiparticle $(-n_e,
-n_m)_{-S}$ simply by $-(n_e, n_m)_S$, one finds \cite{BF}
\begin{eqnarray}
\sw &=& \left\{ \pm (2,0)_0 , \ \pm (1,0)_1 ,\  \pm (2n,1)_{1/2} , 
\pm (2n+1,1)_{-1/2} ,\ n\in \Z \right\} \cr
&{}& \cr
\sso &=&  \left\{ \pm (0,1)_{1/2} , \ \pm (-1,1)_{-1/2} , 
\ \pm (1,0)_{1/2} \right\} \cr
&{}&
\label{xv}
\end{eqnarray}
with states in $\ssp$ or $\ssm$ related to the description in 
$\sso$ by the appropriate
monodromy matrices: $\ssp = \pmatrix{\hfill 2&1\cr -1 & 0\cr} \sso$
and  $\ssm = \pmatrix{1&0\cr 1&1\cr} \sso$. 
One sees that the state $(1,0)_{1/2}$ in $\sso$
corresponds to $(2,-1)_{1/2}$ in $\ssp$ or to $(1,1)_{1/2}$ in 
$\ssm$ and is the one
responsible for the third singularity. Also note that $\sw$ 
contains the W-boson
$(2,0)_0$ and  the quark $(1,0)_1$ as well as dyons with all integer $n_e$. 
All decays across the curve \C\
are consistent with conservation of the mass and of all quantum 
numbers, i.e. electric
and magnetic charges, as well as the $SO(2)$ flavour charge. 
For example, when crossing
\C\ into ${\cal R}_{S0}$, the quark decays as 
$(1,0)_1 \to (0,1)_{1/2}+(1,-1)_{1/2}$.

\vskip 3.mm
\noindent
\underline{$N_f=2$}

\noindent
This case is very similar to the pure Yang-Mills case. The global discrete symmetry
acting on the Coulomb branch of moduli space is again $\Z_2$ and the curve of marginal
stability is exactly the same, cf. Fig. 2, with the singularities again due to a
massless magnetic monopole $(0,1)$ and a massless dyon $(1,1)$. Note however, that this
is in the new normalisation where the W boson is $(2,0)$. So this dyon has half the
electric charge of the W, contrary to what happened for $N_f=0$. With the present
normalisation one finds the weak and strong-coupling spectra as
\begin{eqnarray}
\sw &=& \left\{ \pm (2,0) , \ \pm (1,0) ,\  \pm (n,1) ,\ n\in \Z \right\} \cr
&{}& \cr
\ssp &=&  \left\{ \pm (0,1) , \ \pm (-1,1) \right\}
\label{xvi}
\end{eqnarray}
and all decays across \C\ are again consistent with all quantum numbers. For the quark
one has e.g. $(1,0)\to (0,1)+(1,-1)$ with the flavour representations of $SU(2)\times
SU(2)$ working out as $({\bf 2}, {\bf 2}) = ({\bf 2}, {\bf 1})\otimes ({\bf 1}, {\bf
2})$.

\vskip 3.mm
\noindent
\underline{$N_f=3$}

\noindent
In this case the global symmetry of the action is $\Z_4$ and a given 
vacuum is
invariant under the full $\Z_4$. Consequently, there is no global 
discrete symmetry
acting on the Coulomb branch of the moduli space. There are two singularities
\cite{SWII}, one due to a massless monopole, the other due to a 
massless dyon $(-1,2)$
of {\it magnetic} charge 2. The existence of magnetic charges 
larger than 1 is a
novelty of $N_f=3$. 
The curve of marginal stability again goes through the two singular
points. It is a shifted and rescaled version of the corresponding 
curve for $N_f=0$. Due to the cuts, again we need to introduce two 
different descriptions
of the same strong-coupling BPS state. The variation of $a_D/a$ 
along the curve \C\ is
from $-1$ to $-1/2$ on ${\cal C}^+$ and from $-1/2$ to $0$ on 
${\cal C}^-$. Luckily,
this is such that we do not need any global symmetry to determine 
the strong-coupling
spectrum. For the weak-coupling spectrum, one uses the $M_\infty$ 
symmetry. One finds
\begin{eqnarray}
\sw &=& \left\{ \pm (2,0) , \ \pm (1,0) ,\  
\pm (n,1) , \pm (2n+1,2) , \ n\in \Z \right\} \cr
&{}& \cr
\ssp &=&  \left\{ \pm (1,-1) , \ \pm (-1,2) \right\}
\label{xvii}
\end{eqnarray}
with $(1,-1)\in \ssp$ corresponding to $(0,1)\in \ssm$, so this is really the magnetic
monopole.

The flavour symmetry group is $SU(4)$, and the quark $(1,0)$ is in the representation
${\bf 6}$, the W boson $(2,0)$ and the dyons of magnetic charge two are singlets, while
the dyons $(n,1)$ of magnetic charge one  are in the representation 
${\bf 4}$ if $n$ is even and in
${\overline {\bf 4}}$ if $n$ is odd. Antiparticles are in the complex conjugate
representations of $SU(4)$. Again, all decays across the curve \C\ are consistent with
all quantum numbers, and in particular with the $SU(4)$ Clebsch-Gordan series. As an
example, consider again the decay of the quark, this time as 
$(1,0)\to 2\times (0,1) + (1,-2)$.
The representations on the l.h.s. and r.h.s. are ${\bf 6}$ and 
${\bf 4}\otimes {\bf 4}\otimes {\bf 1}$. Since 
${\bf 4}\otimes {\bf 4}={\bf 6} \oplus {\bf 10}$ this decay
is indeed consistent. All other examples can be found in \cite{BF}.

\def\IM{\mathop{\Im m}\nolimits}
\def\RE{\mathop{\Re e}\nolimits}

\subsection{Massive hypermultiplets, RG flows and superconformal points}

Introducing bare masses for the quark hypermultiplets adds a 
non-negligible technical complication to the previous stability 
analysis: the BPS mass formula now is modified and becomes
\be\label{massivebps}
m_{\rm BPS}(u) =\rd\,
\left\vert n_m \ad(u) -n_e a(u) + \sum_i s_i {m_i\over\rd} 
\right\vert \ ,
\ee
where the $s_{i}$ are  integers or
half-integers which correspond to constant parts of the physical 
baryonic charges \cite{SWII,FERI}. Indeed the fractional fermion numbers 
$S_i(u)$ are non-trivial sections over  the moduli space. While their 
$u$-dependent part already is included in the relevant $a(u)$ and 
$a_D(u)$, for each type of BPS state there is a constant part $s_i$ 
that cannot be consistently removed by shifting the $a(u)$ or 
$a_D(u)$. The bottom line is that there are $N_f$ non-vanishing 
quantum number $s_i$ for each BPS multiplet and they appear in the 
BPS mass formula multiplying the bare masses $m_i$ of the quarks.
This implies that there is not a single curve of marginal stability, 
but an infinity, one for each decay mode.

While this complicates a lot the analysis of the spectra of stable 
BPS states, it also opens the way to studying much richer systems  
like e.g. larger gauge groups. The $m_i$ appear in the BPS mass 
formula as additional parameters, much as the coordinate on moduli 
space $u$, so understanding this case opens the door to studying 
higher rank gauge groups where the moduli space has complex 
dimension larger than one. 

Another phenomenon which can and does occur is the appearance 
of superconformal points. As one varies the masses, the 
singularities move around in moduli space and for certain 
special values of the masses several singularities coincide. 
At these points, several, mutually non-local states have 
vanishing BPS mass and the theory is superconformal.
This will be discussed below.

Again, in this subsection, we only give a rather brief summary 
of this complex situation, and refer the reader to 
\cite{BFmassive} for more details.

\vskip 3.mm
\noindent
\underline{Decay curves}

\noindent
It follows from the above BPS mass formula \eqn{massivebps}
and the same type of reasoning as in the previous section that
a BPS state is stable against any decay of the type
\be\label{qii}
(n_e,n_m)_{s_i} \to k \times (n_e',n_m')_{s_i'} + l \times
(n_e'',n_m'')_{s_i''}
\ee
($k,l \in \Z$) unless  this satisfies at the same time the conservation of
charges and of
the total BPS mass:
\be\label{dqiii}
n_e=k n_e'+l n_e '' \quad , \quad
n_m=k n_m'+l n_m '' \quad , \quad
s_i=k s_i'+l s_i'' \quad \Rightarrow \quad Z=k\, Z'+l\, Z''
\ee
and
\be\label{dqiv}
\vert Z \vert= \vert k\,  Z'\vert + \vert l\,  Z''\vert
\ee
with obvious notations for $Z'$ and $Z''$. If all bare masses $m_i$ are
equal, due to the
${\rm SU}(N_f)$ flavour symmetry,
only the sum $s=\sum_i s_i$ is relevant and needs to be conserved. We
see
that a decay that satisfies the charge conservations \eqn{dqiii} is possible
only if
\be\label{dqv}
{Z'\over Z} \equiv \zeta \in \R \ ,
\ee
and moreover if it is kinematically possible, i.e. if
\be\label{dqvi}
0\le k \zeta \le 1 \ .
\ee
For the case of vanishing bare masses, $m_i=0$ condition \eqn{dqv} reduces to
$\IM {\ad(u)\over a(u)}=0$ which yields a single curve \C$^0$ on the Coulomb
branch
independent of the initial state $(n_e, n_m)_{s_i}$ considered. For
non-vanishing bare
masses however, we have a whole family of possible decay curves. Moreover,
a priori,
there is a different family of such curves for each BPS state. As an
example consider a
dyon with $n_m=1$. Then condition \eqn{dqv} reads
\be\label{dqvii}
\IM\ {n_m' \ad - n_e' a +\sum_i s_i' {m_i\over \rd} \over
\ad - n_e a +\sum_i s_i {m_i\over \rd} } = 0
\ \Leftrightarrow \
\IM\ {-(n_e'-n_m' n_e) a + \sum_i (s_i'-n_m' s_i) {m_i\over \rd} \over
\ad - n_e a +\sum_i s_i {m_i\over \rd} } = 0
\ .
\ee
For fixed $n_e$ and $s_i$, this is an $N_f$-parameter family of curves
with rational parameters
$r_i=(n_e'-n_m' n_e)/(s_i'-n_m' s_i)$. Even though there are some
relations between the
possible quantum numbers $n_e'$ and $s_i',\, n_m'$ 
there are still
many possible values of $r_i$ and we expect a multitude of curves of
marginal stability
on the Coulomb branch of moduli space resulting in a rather chaotic
situation.
Fortunately not all of these curves satisfy the additional criterion
\eqn{dqvi}.
In particular, for
the case of $N_f=2$ with
equal bare
masses, where one expects a different one-parameter family of curves
labelled by
$r=(n_e'-n_m' n_e)/(s'-n_m' s)$, $s=s_1+s_2$, for {\it each} BPS
state, it turned out \cite{BFmassive} 
that only one or two such curves in each family are relevant, i.e.
satisfy the
additional criterion \eqn{dqvi}. Hence the set of all relevant curves 
for {\it all} BPS states
are nicely described by a single set of curves ${\cal C}_{2n}^\pm,\ n\in\Z$, 
and rather
than having a chaotic situation one gets a very clear picture of which
states exist in
which region of the Coulomb branch.

One particularly simple case is the decay of states with $\sum s_i m_i =0$
into states
with $\sum s_i' m_i =0$. The corresponding decay curves all are
given by $\IM {\ad(u)\over a(u)}=0$, i.e. they all coincide with the curve
\C$^0$. This is quite an important case, and  this curve \C$^0$
still plays a priviledged r\^ole, even for non-zero bare masses.

Note that if we had considered decays into three independent BPS states,
$(n_e,n_m)_{s_i} \to k \times (n_e',n_m')_{s_i'} + l \times
(n_e'',n_m'')_{s_i''}
+ q \times (n_e''',n_m''')_{s_i'''}$, we would have {\it two }
conditions: eq. \eqn{dqv} would be
supplemented by ${Z''\over Z}\in \R$, so that such ``triple" decays can
only
occur at the intersection {\it points} of two curves. 
When transporting a BPS state along a path from one region to another, 
the path can
always be chosen so as to avoid such
intersection points. Hence, triple decays  are irrelevant for establishing
the existence domains of the BPS states. Obviously, ``quadruple" and
higher decays, if possible at all, are just as irrelevant.

In order to determine the BPS spectra at
any point on the Coulomb branch, it is most helpful to use 
the following reasonable claim:

\noindent
$\bullet$
{\bf P4}: {\it At any point of the Coulomb branch of a theory having $N_f$ 
flavours with
bare masses $m_j$, $1\leq j\leq N_f$, the set of stable BPS states is included
into the set of stable BPS states of the $m_j=0$ theory at weak coupling.}

Note that the Coulomb branch of the $m_j=0$ theory is separated into
two regions, one containing all the BPS states stable at weak coupling, and
the other at strong coupling containing a finite subset of the BPS states
stable at weak coupling \cite{FB,BF}. One simple consequence of the 
claim (P4)
is that the set of stable BPS states cannot enlarge when one goes from the
$N_f$ to the $N_f -1$ theory following the RG flow 
which is what one naturally expects. This is perfectly
consistent with the spectra determined for zero bare masses in \cite{FB,BF}.
Another consequence, which played a prominent r\^ole in the work 
of ref \cite{BFmassive}, is
that the possible decay reactions between BPS states are then extremely
constrained and thus the number of relevant
curves of marginal stability enormously decreased. 

The detailed analysis is still quite complicated and lengthy and will 
not presented here. We refer instead to the original paper \cite{BFmassive}

\vskip 3.mm
\noindent
\underline{$N_f=2$ with $m_1=m_2\equiv m$}

\noindent
As an example we present the situation for the theory with two 
hypermultiplets, $N_f=2$, having equal bare masses $m_1=m_2=m$. 
There are now 
3 singularities, all on the real axis. We call them $\s_i$ with 
$\s_1\le \s_2\le\s_3$. One has to distinguish two regimes according 
to whether $m$ is smaller or larger than a certain critical value 
which is $\Lambda_2$, $\Lambda_2$ being the relevant dynamically 
generated mass scale. Here, we will focus on $m<\Lambda_2$.

\begin{figure}[htb]
\vspace{9pt}
\centerline{ \fig{12cm}{16cm}{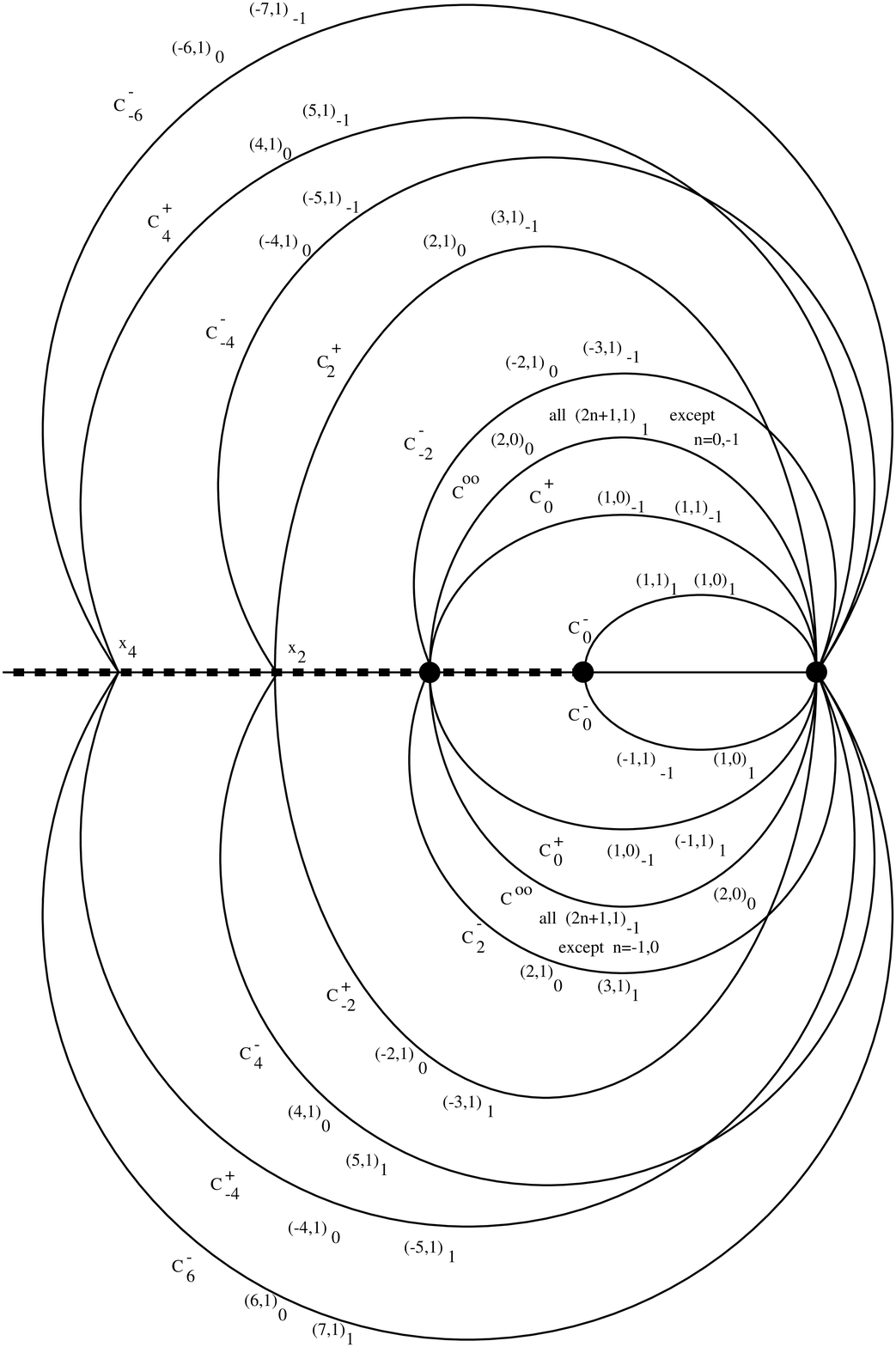} }
\caption{
Shown are a sketch of the relative positions of the relevant decay curves 
for $m<{\Lambda_2\over 2}$ (for not too 
large $\vert n_e\vert$) as well as
the BPS states that decay across these curves. Three states do not decay
anywhere and still are present in the innermost region inside ${\cal
C}_0^-$. They are described as $(0,1)_0$ and $(-1,1)_{\pm 1}$ in the upper, 
and as $(0,1)_0$ and $(1,1)_{\pm 1}$ in the
lower half plane.
Note that, in reality, the angles at which the curves meet the real axis at the
points $x_k$ are slightly different from what they appear to be in
the Figure: indeed, the curves ${\cal C}^-_{-k-2}$, resp.  ${\cal C}^+_{k}$,
in the upper half plane are the smooth continuations of the curves 
${\cal C}^+_{-k}$, resp.  ${\cal C}^-_{k+2}$, in the lower half plane,
in agreement with the monodromy around infinity.
}
\label{massivecurves}
\end{figure}

We have assembled all the {\it relevant} decay curves into 
Fig. \ref{massivecurves} that sketches their
relative positions and indicates the BPS states that decay 
across these curves.
All curves go through $\s_3$, while the other intersection point with 
the real axis depends on the curve: they are $\s_2,\ \s_1$ 
and certain points $x_{2n},\ n=1,2,\ldots$

There are several types of states: first, we have the states that 
become massless at the 
singularities. These are $(0,1)_0$ and, due to the cuts described differently 
in the two half planes, $(0,1)_0$ and $(-1,1)_{\pm 1}$ in the upper, 
and $(0,1)_0$ and $(1,1)_{\pm 1}$ in the
lower half plane. These states exist everywhere (throughout the 
corresponding half plane).

Second, we have the other dyons of $n_e=\pm 1$, the quarks and the W-boson. 
These states
decay on curves in the inner, strong coupling region of the Coulomb 
branch of moduli space:
The W-boson decays on ${\cal C}^\infty$, the quark $(1,0)_{-1}$ on 
${\cal C}^+_0$ and the
quark $(1,0)_1$ on the innermost curve ${\cal C}^-_0$, while the 
dyons $(\e,1)_{-\e}$
decay on ${\cal C}^+_0$ and the dyons $(\e,1)_\e$ on ${\cal C}^-_0$.

Third, we have the dyons with $\vert n_e\vert \ge 2$. 
As discussed above, among these one must distinguish two
sorts: those that will survive the RG flow $m\to\infty$ 
to the pure gauge theory and those
that do not.\footnote{
Note that according to the way $\ad$ and $a$ are defined here, 
the dyons $(2n,1)$ of the $N_f=0$ theory correspond to 
$(2n+1,1)_{\pm 1}$ in the massive $N_f=2$ theory.}
The dyons that will survive this RG
flow are $(2n+1,1)_1$ in the upper half plane and  $(2n+1,1)_{-1}$ in the 
lower half plane. These dyons ($n\ne -1,0$) all decay on the curve 
${\cal C}^\infty$ which thus
plays a priviledged role. The other dyons, namely $(2n,1)_0$ ($n\ne 0$), and 
$(2n+1,1)_{-1}$ in the upper and $(2n+1,1)_1$ in the 
lower half plane ($n\ne -1,0$)
decay on curves ${\cal C}^\pm_{2k},\ k\ne 0$ (where $\vert 2k\vert$ equals 
$\vert n_e\vert$, $\vert n_e\vert+1$ or $\vert n_e\vert-1$).
There are only two states that decay on each of these curves
${\cal C}^\pm_{2k},\ k\ne 0$. These curves move more and more 
outwards as $m$ is increased.
Also, as $\vert k\vert$ gets bigger (i.e. the $\vert n_e\vert$ 
of the corresponding dyons increase) 
these curves more and more reach out towards the semiclassical region.
Conversely, as $m\to 0$, all curves flow towards a single curve, say 
${\cal C}^\infty$.

There are a couple of other points worth mentioning. First remark, that the whole
picture is compatible with the $CP$ transformation 
$(n_e,n_m)_s \to (-n_e,n_m)_{-s}$
under reflection by the real $u$-axis. Second, since all curves go through the
singularity $\s_3$, i.e. all existence domains touch $\s_3$, 
it follows that at this point all BPS
states exist. The same is true for the points $u$ that lie on 
the part of the real $u$ line to
the right of $\s_3$. Indeed, as $\vert n_e\vert$ is increased, 
the corresponding dyon
curves leaving $\s_3$ to the right with an ever smaller slope 
get closer and closer to any
given point on the real interval $(\s_3,\infty)$ but never touch it. 

Finally we note that the whole picture is perfectly consistent: 
if a BPS state decays across a given curve, the decay products are also
BPS states that must exist in the region considered, i.e. on both 
sides of the curve. 
Indeed, this is always the case. As
an example, consider  the dyons $(2n,1)_0$ ($n\ge 1$). In the upper half plane
they decay on the curves ${\cal C}^+_{2n}$ into the dyons $(2n-1,1)_1$ 
and the quark
$(1,0)_{-1}$. These dyons $(2n-1,1)_1$ exist everywhere in the upper half plane 
outside ${\cal C}^\infty$, while the quark $(1,0)_{-1}$ exists 
everywhere outside ${\cal C}^+_0$, and in
particular in the vicinity of the decay curves of $(2n,1)_0$ considered. 

\vskip 5.mm
\noindent
\underline{Superconformal points}

\noindent
As the mass $m$ is increased and gets closer to $\Lambda_2/2$, the 
singularities $\s_2$ and $\s_3$ approach each other and eventually 
coincide at $m=\Lambda_2/2$. There we have a superconformal point. 
As $m$ is increased beyond, the singularities separate again but 
correspond to different states that become massless there. The 
analysis of the decay curves and stable states then is different 
from, but analogous to the case  $m<\Lambda_2/2$, and we refer 
the reader to \cite{BFmassive}. Let us now look at $m=\Lambda_2/2$.

Call $M_i$ the monodromy matrices around $\s_i$ for  $m<\Lambda_2/2$, 
and $M'_i$ the monodromy matrices around $\s_i$ for  $m>\Lambda_2/2$.
Clearly, $M_1=M'_1$ since the singularity $\s_1$ is not affected 
by the collision of $\s_2$ and $\s_3$. Also the product of the 
monodromies around $\s_2$ and $\s_3$  should not be affected. Such 
a statement however needs to be made with care since the precise 
definition of the monodromy matrices depends on the analytic structure, 
i.e. how one arranges the different cuts along the real axis. With an 
appropriate convention however, one has
\be\label{scmono}
M_{\rm sc}= M_3 M_2 = M'_3 M'_2 \ .
\ee
$M_{\rm sc}$ then is the monodromy around the collapsed singularity 
$\s_2=\s_3$ at the superconformal point. This monodromy matrix is
\be\label{monosmatrix}
M_{\rm sc}=\pmatrix{ 0&1\cr -1&0\cr} \equiv - S \in SL(2,\Z) \ ,
\ee
so that $S$-duality must be a symmetry at the superconformal point.

Which are the massless states at the superconformal point? For
$m<\Lambda_2/2$ the massless states are $(-1,1)_{-1}$ and $(1,1)_1$ 
at $\s_2$ and $\s_3$ respectively. For
$m>\Lambda_2/2$ the massless states are $(0,1)_0$ and $(1,0)_1$ at 
$\s_2$ and $\s_3$ respectively. In ech case the massless states at 
$\s_2$ and $\s_3$ are exchanged by $M_{\rm sc}=-S$.
At the superconformal point one might expect to have two massless states, 
but then the question would be which ones. It turns out that actually all of 
these states, $(-1,1)_{-1},\ (1,1)_1,\ (0,1)_0$ and $(1,0)_1$ 
exist at the superconformal point and are massless. Furthermore, all other BPS states also exist at this point, but are heavy.

One can extract quite a lot of information about the superconformal theory 
simply from studying the monodromy matrices. We refer the interested reader 
to \cite{BFmassive}.
%



\end{document}